\documentclass[11pt,a4paper]{article}
\usepackage[utf8]{inputenc}
\usepackage[english]{babel}
\usepackage[T1]{fontenc}
\usepackage{natbib}

\usepackage{amsfonts}
\usepackage{amsthm}
\usepackage{amsmath}
\usepackage{amssymb}
\usepackage{mathrsfs}
\usepackage{mathtools}
\usepackage[toc,page]{appendix}
\usepackage{dsfont} 
\usepackage{enumerate}
\usepackage{graphicx}
\usepackage{color}

\usepackage{float}
\usepackage{makecell}
\usepackage{times}
\usepackage{lineno} 
\usepackage{fancyhdr}
\usepackage[ruled,vlined]{algorithm2e}
\usepackage{caption}
\usepackage{subcaption}
\usepackage{tkz-graph}
\usepackage{tikz}
\tikzstyle{int}=[draw, line width = 0.5mm, minimum size=5em]
\usetikzlibrary{positioning,automata,arrows}
\usetikzlibrary{matrix}
\usepackage{soul} 
\setstcolor{red}
\usepackage[colorlinks=true,allcolors=blue]{hyperref}

\usepackage{algpseudocode}
\usepackage[normalem]{ulem} 
\newcommand{\RR}{\textsf{R }} 

\def\modif#1{{\color{black} #1}} %

\usepackage{soul}
\setstcolor{blue}

\author{Henri Mermoz KOUYE$^{(1,2)}$, Gildas MAZO$^{(1)}$, Cl\'ementine PRIEUR$^{(2)}$ \\
and Elisabeta VERGU$^{(1,\dagger)}$\\
 \footnotesize{$^{(1)}$Univ.  Paris-Saclay, INRAE, MaIAGE, 78350, Jouy-en-Josas, France}\\
 \footnotesize{$^{(2)}$Univ. Grenoble Alpes, CNRS, Inria, Grenoble INP, LJK, 38000, Grenoble, France}\\
  \footnotesize{$^{(\dagger)}$E. Vergu died on May 13, 2023, during the writing of this paper}}
\title{\modif{Performing global sensitivity analysis on simulations of a continuous-time Markov chain model motivated by epidemiology}}

\newtheorem{proposition}{Proposition}

\newtheorem{example}{Example}

\newcommand{\set}[1]{\mathbf{#1}}

\newcommand{\X}{\mathbf{X}}

\usepackage{mwe}
\usepackage{comment}

\newlength{\tempdima}
\newlength{\tempheight}
\newlength{\tempwidth}
\newcommand{\rowname}[1]
{\rotatebox{90}{\makebox[\tempdima][c]{\textbf{#1}}}}
\newcommand{\columnname}[1]
{\makebox[\tempwidth][c]{\textbf{#1}}}

\begin{document}

\maketitle
\begin{abstract}
\modif{In this paper we apply a methodology introduced in \cite{Navarro} in the framework of chemical reaction networks to perform a global sensitivity analysis on simulations of a continuous-time Markov chain model motivated by epidemiology. Our goal is to quantify not only the effects of uncertain parameters such as epidemic parameters (transmission rate,
  mean sojourn duration in compartments), but also
  those of intrinsic randomness and interactions between epidemic parameters and intrinsic randomness. For that purpose, following what was proposed in \cite{Navarro}, we leverage three exact simulation algorithms for continuous-time Markov chains from the state of the art which we combine with common tools from variance-based sensitivity analysis as introduced in \cite{sobol}. Also, we discuss the impact of the choice of the simulation algorithm used for the simulations on the results of sensitivity analysis. Such a discussion is new, at least to our knowledge. In a numerical section, we implement and compare three sensitivity analyses based on simulations obtained from different exact simulation algorithms of a SARS-CoV-2 epidemic model.}
\end{abstract}


\section{Introduction}
In epidemiology, stochastic compartmental models \modif{help in} the
prediction and understanding of spreads of infectious diseases in a
host population, such as humans, animals, or plants. The \modif{output} of
those models \modif{usually depends} on numerous uncertain parameters such as
transmission rate, \modif{mean infectious period} or mean sojourn duration \modif{in  the different compartments}. The aim of sensitivity analysis is to
identify, among these parameters, the ones which have the greater
impact on the infection
spread~\citep{hanthanan2023exploring,goel2023sirc,MASSARD2022111117}.  This is useful,
e.g., \modif{for elaborating} efficient control
strategies~\citep{ngonghala2015persistent,yang2016effect} or \modif{performing model comparisons}~\citep{torii2023global}.  In the following, we \modif{focus on} global sensitivity analysis (GSA) of
stochastic compartmental models \modif{commonly} used in epidemiology~\citep{courcoul}. We focus \modif{on GSA} rather than local sensitivity analysis, as the former is better
adapted to nonlinear models.

Compartmental models \citep{Brauer2008} consist in dividing the host population into compartments, each containing individuals with a similar health status. \modif{Health status of each individual changes over time.} 
\modif{Transitions between compartments are highly dependent on individual characteristics or the contact pattern between individuals.}
While in large populations randomness due to individual-to-individual variability averages out, it has a large impact on the transmission process for small populations \citep{Britton2009StochasticEM,bittihn_stochastic_2020}. As they incorporate stochastic effects related to biological or contact events, 
stochastic compartmental models are used to analyze thoroughly the outbreak of infectious diseases. Throughout \modif{this paper}, we thus focus on stochastic models, and more precisely on continuous-time Markov chains (CTMC). \modif{A CTMC is a continuous time and memoryless discrete event stochastic process, which means that the past history impacts on the future evolution of the system only via the current state of the system. For a CTMC, inter-event times are exponentially distributed.}

GSA aims at determining the extent to which the variability of an input parameter or of a set of input parameters affects the variability of \modif{model output} (see, e.g., \cite{SAL2000,MARINO2008178}). \modif{In the framewrok of a deterministic model, a
variance-based global sensitivity analysis is often used to rank the importance of input parameters (or sets of input parameters), based on their contribution to the variance of the output quantity of interest (QoI), via the computation of the so-called Sobol' indices introduced in \cite{sobol}. Performing GSA for stochastic models is more complex as the model output is tainted with two sources of uncertainty: the intrinsic randomness of the model and the uncertainty on \modif{epidemic} model parameters (such as mean sojourn duration in \modif{the different compartments}, transmission rate and others).
  So far, different paradigms have been introduced in the \modif{litterature} for sensitivity analysis of stochastic models.}
  
\modif{A pragmatic approach for GSA of stochastic models consists in performing the analysis on both expectation and variance of model output conditionally on epidemic uncertain parameters.} Intrinsic randomness is thereby considered as noise and is smoothed by \modif{averaging}. \modif{More generally, this approach amounts to summarizing the output of the stochastic model by several deterministic QoIs (obtained by smooting intrinsic random noise) and then to perform a GSA for each of these QoIs, e.g., by computing Sobol' indices.} This approach is often used in practice in various applications, for instance in: \cite{courcoul} to identify key parameters of a model describing the spread of an animal disease in a cattle herd; \cite{Rimbaud} for a model describing the spatio-temporal spread of plant pathogens; \cite{Richard} for a SARS-CoV-2 spread model; \cite{Cristancho} for a theoretical metapopulation model. 
\modif{It is important to note that if we evaluate a stochastic model multiple times for the same values of uncertain input parameters, we obtain different values for the model output, due to the intrinsic randomness of the model. In this framework, putting in practice the aforementioned approach relies on a fine trade-off between exploration of parameter space and repetition} \citep{mazo}. To avoid computational burdens \modif{due to the cost of model evaluations, it is possible to perform GSA on a metamodel}. \modif{A metamodel/ surrogate model is a simplified model approximating the actual model at a much lower cost.}
In \cite{marrel_global_2012}, the mean and variance are \modif{jointly emulated by Gaussian process regression (see, e.g., \cite{williams2006gaussian}). In \cite{etore}, the sensitivity analysis of certain QoIs calculated from a stochastic differential equation, for example a hitting time, is based on metamodeling by polynomial chaos expansion.}
\modif{The aforementioned approach, based on deterministic QoIs to summarize the results of a stochastic model, suffers from two main drawbacks. The first is that certain parameters may prove influential on a QoI but not on others, because the GSA is carried out separately on the different QOIs. The second is that this approach provides no information on how intrinsic noise influences the model output.} 

More recently, a different point of view was adopted (see, e.g., \cite{fort2020global,daveiga}). Stochastic models \modif{are} interpreted as deterministic models \modif{whose output for a given set of model parameter values is a probability distribution. Then it is possible to emulate the output probability distribution (see, e.g., \cite{zhu, Zhu2023})  and/or to define sensitivity indices that measure the sensitivity of the output probability distribution to variations of input parameters.} 

\modif{In the framework we consider in the present paper, it is possible to control intrinsic randomness (e.g., by fixing the seed in a code). It is the framework considered in \cite{Hart} where Sobol' indices are computed for each realization of the variable controlling the internal noise. It results in random Sobol' indices, whose randomness is inherited from the intrinsic randomness of the model. Still in this framework, a different approach is adopted
  in~\cite{LEMAITRE2015107,jimenez,Navarro}. It is the one we focus on in our work. The stochastic algorithm
  used to simulate the model output is reinterpreted as a
  deterministic one with an augmented set of inputs comprising both
  the uncertain parameters and the latent variables controlling
  intrinsic randomness. This allows to apply standard GSA tools to the
  algorithm defined on the augmented input space, and, thereby,
  quantify not only the effects of the uncertain parameters (the epidemic parameters in our setting),
  but also those of intrinsic randomness and the interaction effects
  between the uncertain parameters and intrinsic
  randomness. We shall call this the complete GSA method.}

\modif{In~\cite{LEMAITRE2015107,jimenez,Navarro}, the Modified Next
  Reaction Method algorithm, which is well-known in the field of simulation of chemical reaction networks, was used to simulate the Markov
  chain. However, in epidemiology, it is customary to use Gillespie
  algorithms. One may then ask whether it is possible to apply the
  complete GSA method based on these algorithms, and wether this
  would lead to similar results for sensitivity analysis as intrinsic randomness is modeled differently from one algorithm to the other. To address this
  problem, we first review three exact stochastic simulation algorithms---namely,
  Gillespie Direct Method~\citep{GILLESPIE}, Gillespie First
  Reaction Method~\citep{GILLESPIE} and Modified Next Reaction
  Method~\citep{anderson}---and then apply the complete GSA method
  on each of these algorithms. Our contribution is to
  show, both mathematically and numerically, that the results of the variance-based sensitivity analysis depend on the chosen algorithm. To the best of
  our knowledge, this point has never been discussed in the literature so far.}

The paper is organized as follows. \modif{We provide in Section \ref{compart} a description of the class of compartmental models we are interested in, whose mathematical formulation is a CTMC. In Section \ref{gsa} we recall the definition of variance-based Sobol' indices \citep{sobol} for sensitivity analysis of deterministic models. In Section \ref{sto_framework} we review the methodology of global sensitivity analysis of a stochastic model consisting in representing this model as a deterministic model with input space the set of epidemic parameters augmented with the set of latent variables modeling intrinsic randomness. In Section \ref{SAfree}, we exhibit a toy example showing that such a representation is not unique and that GSA results depend on it. Thus GSA results have to be interpreted with caution.
In Section \ref{secrepr}, we review the most common exact simulation algorithms for CTMC stochastic compartmental models and 
for each of them we provide the corresponding deterministic algorithm used for a complete GSA method. We illustrate  our approach in Section \ref{covid} by considering 
a parsimonious SARS-CoV-2 spread
model as a case study. We compare and discuss the GSA results obtained with the different representations presented in Section \ref{secrepr}. Finally the main conclusions of our study are recalled in Section \ref{conclu}.}

\section{Preliminaries}
\label{sec:GsaForStoCompModels}
We first recall in Section \ref{compart} the definition of CTMC stochastic compartmental models we are interested in. Then in Section \ref{gsa} we recall the definition of variance-based Sobol' indices \citep{sobol} used for GSA in the framework of deterministic models.


\subsection{CTMC stochastic compartmental models}
\label{compart}

Consider a finite, closed (i.e. of constant size over time) population in which each individual has a
health status (susceptible, infectious, and so on) evolving over
time. The set of all possible health statuses is denoted by $\set
V$. Since those health statuses induce a partition of the whole
population at any given time, the elements of $\set V$ are also called
compartments. Every time an individual changes compartments, we say
that a transition occurs. Only certain types of transitions can
occur. Let $\set E$ denote the set of all possible types of
transitions. By definition, for $\alpha,\beta\in\set V$, an individual
can move from $\alpha$ to $\beta$ if $(\alpha,\beta)\in\set E$. The
pair $(\set V,\set E)$ can be identified with a directed graph where
$\set V$ is the set of nodes and $\set E$ is the set of arrows
connecting the compartments between which the individuals can
move. Assuming an ordering of the compartments has been chosen, let us
identify the set $\set E$ with a subset of $\mathbb{R}^{|\set V|}$ as
follows: with each $(\alpha,\beta)\in\set E$, associate the vector of
length $|\set V|$ with components equal to zero except at the
positions corresponding to $\alpha$ and $\beta$, where the components
are $-1$ and $1$, respectively. The elements of $\set E$ seen as a
subset of $\mathbb{R}^{|\set V|}$ are called transition vectors.

For every $\alpha\in\set V$ and a vector of epidemic parameters
$\theta\in\Theta\subset\mathbb{R}^d$, let $W^{\theta}_{\alpha}(t)$ be
the number of individuals in compartment $\alpha$ at time $t$. Since
the population is closed, we have that
$\sum_{\alpha\in\set V}W^{\theta}_{\alpha}(t)$ is constant over
time. Denote by $W^{\theta}=(W_{\alpha}^{\theta})_{\alpha\in\set V}$
the stochastic process that describes the whole population over time.
It is assumed that $W^\theta$ is a continuous-time Markov chain with
state space $\mathcal{E}\subset \mathbb{N}^{|\set{V}|}$ and
positive rate functions $g_{\mathbf{u}}(\theta,\xi)$, $\mathbf{u}\in\set E$,
$\xi\in\mathcal{E}$, given by
\begin{linenomath}
\begin{equation*}
  \mathrm{Pr}(W^{\theta}(t+s)=\xi+\mathbf{u}|W^{\theta}(t)=\xi)
  = g_{\mathbf{u}}(\theta,\xi)s+o(s) \textup{ as } s\rightarrow 0.
\end{equation*}
\end{linenomath}
The initial state $W^\theta(0)=:\xi_0\in \mathcal{E}$ is supposed to
be fixed.
A description of the CTMC of the classical SIR model is given in Example \ref{exple:sir-process}.
 
\begin{example} The classical SIR model is described as follows.
  There are three compartments $\set{V}=\{S,I,R\}$ and two types of
  transitions: infection $(S,I)$ and removal $(I,R)$ so that
  $\set{E}=\{(S,I),(I,R)\}=\{(-1,1,0),(0,-1,1)\}$.  Infection is
  characterized by the transition vector $\mathbf{u}_{S,I}=(-1,+1,0)$ and
  the rate function $g_{S,I}=\frac{\beta}{N}W_IW_S$, where $\beta$
  is some parameter and $N$ is the total size of the population. Removal has transition
  vector $\mathbf{u}_{I,R}=(0,-1,+1)$ and rate function
  $g_{I,R}=\gamma W_I$, where $\gamma$ is some parameter. The vector
  of parameters is $\theta=(\beta,\gamma)$. The graph of the SIR model is given below:
  \begin{figure}[H]
    \centering
    \resizebox{6cm}{1cm}{
      \begin{tikzpicture}[node distance=1cm,auto,>=latex']
        \node[int] (c) [] {$\mathbf{S}$};

        \node [int] (d) [right of=c, node distance=4cm] {$\mathbf{I}$};

        \node [int] (e) [right of=d, node distance=4cm] {$\mathbf{R}$};

        \draw[->, ultra thick, black] (c) edge node {$\frac{\beta}{N}W_IW_S$} (d);
        \draw[->, ultra thick, black] (d) edge node {$\gamma_IW_I$} (e);
      \end{tikzpicture}}
  \end{figure}
  \label{exple:sir-process}
\end{example}

\subsection{Global Sensitivity Analysis}
\label{gsa}
In this section, we first recall the definition of variance-based Sobol' sensitivity indices introduced in \cite{sobol} for deterministic models with scalar output. 
Following the paradigm of GSA, we model uncertain inputs by a random vector of independent components $\mathbf{X}=(X_1,\dots,X_m)$. Let $E_1,\dots,E_m$ be subsets of $\mathbb{R}$ and $f_1 \, : \, E_1 \times \ldots \times E_m \rightarrow \mathbb R$ be some function such that $\mathbb{E}\left(f_1(\mathbf{X})^2\right) <+\infty$. 
Then first-order and total Sobol' indices (see, e.g., \cite{sobol,HOMMA}) of the ouput $Y=f_1(\mathbf{X})$ associated with input $X_j, j=1,\cdots ,m$, are respectively defined as:
\begin{linenomath}
\begin{align}
S_{X_j} &=\frac{\text{Var}\left(\mathbb{E}\left[Y\mid X_j\right]\right)}{\text{Var}\left(Y\right)}\, ,\label{scal1}\\
S_{X_j}^{\textup{tot}}  &=1-\frac{\text{Var}\left(\mathbb{E}\left[Y\mid X_1,\cdots ,X_{j-1},X_{j+1},\cdots ,X_{m}\right]\right)}{\text{Var}\left(Y\right)} \nonumber\\
 & =:1-\frac{\text{Var}\left(\mathbb{E}\left[Y\mid \X_{-j}\right]\right)}{\text{Var}\left(Y\right)} \, \cdot \label{scal2}
\end{align} 
\end{linenomath}

The definition of first-order and total Sobol' indices can be extended
to models with vectorial or functional output (see, e.g.,
\cite{lamboni,gamboa2}). Let $\modif{Y=(Y_1, \ldots , Y_p)}:=f_2(X_1, \ldots ,
X_m)$ be a vectorial output where $f_2 \, : \, \, E_1 \times \ldots \times E_m \rightarrow \mathbb{R}^p$ is some function and $\mathbb{E}\left(\|Y\|^2\right) <+\infty$, with $\|\cdot\|$ denoting the Euclidean norm on $\mathbb R^p$. Aggregated first-order and total Sobol' indices are defined as:
\begin{linenomath}
\begin{align}
S_{X_j}&=\frac{\sum_{k=1}^p\text{Var}\left(Y_k\right)S_{X_j,k}}{\sum_{k=1}^p\text{Var}\left(Y_k\right)}\, , \label{vec1}\\
S_{X_j}^{\textup{tot}} &=\frac{\sum_{k=1}^p\text{Var}\left(Y_k\right)S_{X_j,k}^{\textup{tot}}}{\sum_{k=1}^p\text{Var}\left(Y_k\right)}\, , \label{vec2} 
\end{align}
\end{linenomath}
where $S_{X_j,k}$ and $S_{X_j,k}^{\textup{tot}}$ are the first-order and total Sobol' indices of the scalar output $Y_k$ associated with the input $X_j$, for $k=1, \ldots , p$ and $j=1, \ldots , m$.
If the output of the model of interest is a function of time, it can be reduced to a vectorial output through discretization of time. Then aggregated first-order and total Sobol' indices can be computed using \eqref{vec1} and \eqref{vec2}, where the output components $Y_k$ would be the values of the function at the time points of the discretization. Also first-order and total indices defined in Equations \eqref{scal1} and \eqref{scal2} can be computed at each time point of the discretization in order to obtain  dynamics of Sobol' indices.

\section{Complete GSA for stochastic models}
\label{sto_framework}

\modif{In this section, we review the methodology of global sensitivity analysis of a stochastic model consisting in representing this model as a deterministic model with input space the set of uncertain parameters augmented with the set of latent variables modeling intrinsic randomness. This methodology permits a quantification of the sensitivity of model output to a variation  of the uncertain parameters but also of latent variables modeling intrinsic randomness, and finally of the interaction
between both. In the following, we call this a
complete GSA. A strategy to achieve this aim requires controlling the latent
variables modeling intrinsic randomness.}

\subsection{Deterministic representations of stochastic models}
\label{sec:detReprOfStoMod}

Let $Y^\theta$ denote the random output of some stochastic model with
parameters $\theta\in\Theta\subset\mathbb{R}^d$. For instance,
$Y^\theta$ might be the stochastic process $W^\theta$ introduced in
Section~\ref{compart}, or any scalar (or vectorial) quantity of
interest defined as a functional of the process $W^\theta$. Note that
distinct values of the parameters encoded in $\theta$ correspond to
distinct epidemiological patterns. We thus consider the collection
$\{Y^\theta \, , \; \theta\in \Theta\}$ and assume mutual \modif{independence}
between its members.

As in Section~\ref{gsa}, uncertain parameters are modeled by a
random vector $\mathbf{X}=(X_1,\dots,X_d)$ with independent
components. In addition, it is assumed that $\mathbf{X}$ is independent of
$\{Y^\theta \, , \; \theta \in \Theta\}$. The pair $(\mathbf{X},Y^{\mathbf{X}})$ represents the
input/output pair that an external observer would see should they
draw  uncertain parameters at random. The object $Y^\theta$ is then
the output observed conditionally on $\mathbf{X}=\theta$.

It is important to note that in $Y^{\mathbf{X}}$ there are two
sources of variability (and hence uncertainty): the one coming from
the uncertainty of the parameters (that is, modeled by vector
$\mathbf{X}$), and the one coming from the intrinsic randomness of
the stochastic model (that is, for a fixed $\theta$ the variability in $Y^{\theta}$). A
complete GSA aims at separating these two sources of uncertainty
and quantifying interactions between both.

To achieve this aim, it is necessary to control the latent variables modeling
intrinsic randomness. More precisely, we aim at finding a function $f$
and a latent or a set of latent variables $Z$, independent of
$\mathbf{X}$, such that the probability distributions of $Y^\theta$
and $f(\theta,Z)$ coincide.  Since $\mathbf{X}$ is independent of $Z$
and $Y^\theta$, we immediately have that the input/output pairs
$(\mathbf{X},Y^\mathbf{X})$ and $(\mathbf{X},f(\mathbf{X},Z))$ are
equal in distribution.  The pair $(f,Z)$ is called a \emph{deterministic
  representation} (or simply a representation) of the
stochastic model $Y^\theta$.  Often, the function $f$ is the
function induced by a (deterministic) algorithm which, if the inputs
of that algorithm were drawn from the right distribution, would
produce an output statistically equal to the given stochastic model.

From the viewpoint of GSA, one advantage of constructing a
deterministic representation of a stochastic model is that standard
methods of GSA for deterministic models can be applied
straightforwardly. For instance, we can compute first-order and total
Sobol' indices by letting $m=d+1$ and $X_{d+1}=Z$ in~\eqref{scal1} and
\eqref{scal2} (or in~\eqref{vec1} and~\eqref{vec2} if the output is
vectorial or functional).

In general, there is no unique deterministic representation of a
stochastic model. The set of latent variables modeling
intrinsic randomness $Z$ and the function $f$ may vary from one representation to the
other. More precisely, 
if $Y^\theta$ is a stochastic model and $(f,Z)$ a
representation of it, there may exist another representation $(\tilde f,\tilde Z)$
of $Y^\theta$ such that the laws of $(\mathbf{X},\tilde f(\mathbf{X},\tilde Z))$
and $(\mathbf{X},f(\mathbf{X},Z))$ coincide. (Here the probability distribution of $\tilde Z$ may
be different from that of $Z$.) An example of a toy stochastic model
 with two different
representations is provided in Example~\ref{example_1} below.

\begin{example}
\label{example_1}
Let $Z\sim \mathcal{U}\left([0,1]\right)$ independent of
$\left(\X,Z_1,Z_2\right)\sim
\mathcal{N}\left(0_{\mathbb{R}^3},\mathrm{Id}_3\right)$. Consider the
stochastic model $Y^\theta\sim\mathcal{N}(\theta,1)$. This model can
be represented by using
$f(\theta,Z_1,Z_2)=\theta+\frac{1}{\sqrt{2}}\left(Z_1+Z_2\right)$ or
$\tilde f (\theta,Z)=\theta+\Phi^{-1}(Z)$, where $\Phi$ is the
cumulative distribution function of the standard normal distribution.
\end{example}

\subsection{How does a complete GSA depend on the chosen representation?}\label{SAfree}

As different representations can be exhibited for a same stochastic
model, we can wonder how the choice of representations affects GSA
results.  Let us consider $(f,Z)$ and $(\tilde f,\tilde Z)$ two
distinct representations of a same stochastic model with uncertain
parameters $\X=\left(X_1,\cdots ,X_d\right)$ and output $Y^{\X}$.
We say that an index $S_{X_j}$ is \emph{representation free} if $S_{X_j}(f,Z)=S_{X_j}(\tilde f, \tilde Z)$, where here $S_{X_j}(f,Z)$ and $S_{X_j}(\tilde f,\tilde Z)$ denote the values of the index $S_{X_j}$ based on the representations $(f,Z)$ and $(\tilde f,\tilde Z)$, respectively.

\begin{proposition}
  First-order Sobol' indices associated with uncertain parameters are representation free.
  \begin{proof}
    From Section \ref{sto_framework} we know that
    \begin{equation}\label{proprepr}
      \left(\X,Y^{\X}\right)\sim \left(\X,f(\X,Z)\right) 
      \sim \left(\X,\tilde f (\X,\tilde Z )\right). 
    \end{equation}
    Thus, for $j=1, \ldots, d$, we have almost surely the following  equality:
    $$\mathbb{E}\left[f(\X,Z)\mid X_j\right]=\mathbb{E}\left[\tilde f (\X,\tilde Z)\mid X_j\right]$$ from which we deduce, using \eqref{scal1}, that $S_{X_j}(f,Z)=S_{X_j}(\tilde f,\tilde Z)$, where here $S_{X_j}(f,Z)$ and $S_{X_j}(\tilde f,\tilde Z)$ denote the first-order Sobol' indices associated with $X_j$ based on $(f,Z)$ and $(\tilde f,\tilde Z)$, respectively.
  \end{proof}
\end{proposition}

\begin{proposition}
  Total Sobol' indices associated with intrinsic randomness are representation free.
  \begin{proof}
    Put $X'_j=X_j$, $j=1,\dots,d$, $X'_{d+1}=Z$ and $m=d+1$ so that $\mathbf{X}':=(X_1',\dots,X_m')=(\mathbf{X},Z)$.
    From \eqref{proprepr}, we deduce 
    the following almost sure equality:
    $$\mathbb{E}\left[f(\X,Z)\mid \mathbf{X}'_{-m}\right]=\mathbb{E}\left[f(\X,Z)\mid \X\right]
    =\mathbb{E}\left[\tilde f (\X,\tilde Z)\mid \X\right]=\mathbb{E}\left[\tilde f(\X,\tilde Z)\mid \mathbf{X}'_{-m}\right],$$
    from which we deduce, using \eqref{scal2}, that the total Sobol' indices $S_{Z}^{\textup{tot}}(f,Z)$ and $S_{\tilde Z}^{\textup{tot}}(\tilde f,\tilde Z)$ associated with intrinsic randomness and based respectively on $(f,Z)$ and $(\tilde f,\tilde Z)$ are equal.
  \end{proof}
\end{proposition}

\begin{proposition}\label{prop:depend}
  First-order Sobol' indices associated with intrinsic randomness and total Sobol' indices associated with uncertain parameters depend on the choice of representations in general.
\end{proposition}

To show that Proposition~\ref{prop:depend} is true, it suffices to exhibit an example where two distinct representations lead to distinct first-order Sobol' indices associated with intrinsic randomness and distinct total Sobol' indices associated with uncertain parameters. Before giving the example, let us give some intuition behind Proposition~\ref{prop:depend}. Note that the random variables  $\displaystyle \mathbb{E}\left[f(\X,Z)\mid (\mathbf{X}_{-j},Z)\right]$ and $\displaystyle \mathbb{E}\left[\tilde f(\X,\tilde Z)\mid (\mathbf{X}_{-j},\tilde Z)\right]$  have different probability distributions in general. Indeed, since $(f,Z)\neq (\tilde f,\tilde Z)$, the way each function $f$ or $\tilde f$ combines its (set of) latent variable(s) $Z$ with input $X_j$ to generate the output may differ. Thus there is no reason for total Sobol' indices associated with uncertain parameters to be representation free. Also, there is no reason for the random variables $\mathbb{E}\left[f(\X,Z)\mid Z\right]$ and $\mathbb{E}\left[\tilde f(\X,\tilde Z)\mid \tilde Z\right]$ to have the same probability distributions and hence the first-order Sobol' index associated with intrinsic randomness to be representation free. This is illustrated on the toy Example \ref{example_2} below.

\begin{example}
\label{example_2}
Let $X$ be a random variable independent of $Z$ and $\tilde Z$ where $Z$ and $\tilde Z$ are i.i.d.  under $ \mathcal{N}(0,1)$. Define two functions: $f(X,Z)=XZ$ and $\tilde f(X,\tilde Z)=X^2 \tilde Z$. If $X$ is distributed such that $\mathbb{P}\left(X=-1\right)=\mathbb{P}\left(X=1\right)=\frac{1}{2}$ then $(X,f(X,Z))\sim (X,\tilde f(X,\tilde Z))$. Thus, $(f,Z)$ and $(\tilde f,\tilde Z)$ represent the same stochastic model but:
$
\mathbb{E}\left[f(X,Z)\mid Z\right]=0$ while $\mathbb{E}\left[\tilde f(X,\tilde Z)\mid \tilde Z\right]=\tilde Z.
$
Simple calculations then lead to
$\displaystyle S_Z(f,Z)=0$ while $\displaystyle S_{\tilde Z}(\tilde f,\tilde Z)=1$. Total Sobol' indices associated to $X$ can easily be deduced: $S_X^{\textup{tot}}(f,Z)=1-S_Z(f,Z)=1$ and $S_X^{\textup{tot}}(\tilde f,\tilde Z)=1-S_{\tilde Z}(\tilde f,\tilde Z)=0$. 
\end{example}

We conclude from this section that intrinsic randomness can be modeled in different manners. GSA results naturally depend on the modeling choice. Different modelings bring different insights. 
We discuss this point further on a SARS-CoV-2 spread model in Section \ref{covid}. In the following section, we exhibit different meaningful representations for CTMC stochastic compartmental models, based on different simulation algorithms.

\section{Deterministic representations for CTMC stochastic compartmental models}\label{secrepr}
\label{representation}

Following Section~\ref{compart}, let $W^\theta$ be a CTMC stochastic
compartmental model with uncertain parameters $\theta$. As explained
in Section~\ref{sto_framework}, we wish to rewrite the trajectories of
$W^\theta$ as a function of the uncertain parameters and some set of
latent variables so as to perform a complete GSA. Based on three
different exact simulation algorithms \modif{from the state of the art}, we propose in this section
three different deterministic representations for the same generic
CTMC stochastic compartmental model.

One of the first and most basic procedure to simulate \modif{trajectories}
of a CTMC is as follows.  Given that the chain is at some state
$W^{\theta}(s)=\xi$ at time $s$, the holding time until the next jump
is distributed as an exponential random variable with parameter
$\sum_{\mathbf{u}\in \set{E}}g_{\mathbf{u}}(\theta,\xi)$ and then the
chain moves to state $\xi+\mathbf{u}$ with probability
$g_{\mathbf{u}}(\theta,\xi)/\sum_{\mathbf{u}\in
  \set{E}}g_{\mathbf{u}}(\theta,\xi)$. See,
e.g.~\cite{KarlinTaylorSecondCourse} for more details. In
epidemiology, this procedure is known as Gillespie Direct Method~\citep{GILLESPIE}.

Algorithm \ref{algo_gil1} is a \modif{slight} modification of Gillespie Direct Method based on two pseudo-random number generators $RG_1$, $RG_2$. \modif{The first is used to find when the next
transition occurs and the second is used to determine which
type of transitions occurs at that time.} Each pseudo-random number generator is seen as an infinite sequence of pseudo-random numbers determined \modif{by a positive integer called the seed of the generator}. \modif{For ${\bf z}=(z_1,z_2)$ a realization of the random seed ${\bf Z}=(Z_1,Z_2)$, we denote by $f(\theta,{\bf z})$ the output of Algorithm~\ref{algo_gil1}. Then $\left(f,{\bf Z}\right)$ is a deterministic representation of $W^{\theta}$ in the sense of Section \ref{sec:detReprOfStoMod}. In practice, random seeds $Z_1$ and $Z_2$ are drawn independently from a uniform distribution over a large set of positive integers.}\\

\begin{algorithm}[H]
\SetAlgoLined
\SetKwInOut{Input}{Inputs}
\SetKwInOut{Output}{Output}
\Input{$t_{\mathrm{end}}$, $\theta$, $Z:=\left(RG_1,RG_2\right)$}
 \KwData{$\xi_0$, $\set{E}$, $\{g_{\mathbf{u}},\mathbf{u}\in \set{E}\}$}
\Output{$\{W^{\theta}(s), s\in [0,t_{\mathrm{end}}]\}$ }
 Initialization: 
 $s\gets 0$, $W^{\theta}(s) \gets \xi_0$\;
 \While{$s < t_{\mathrm{end}}$}{
  
 $\Sigma \gets \sum_{\mathbf{u}\in \set{E}}g_\mathbf{u}\left(\theta,W^\theta(s)\right)$\;
 Take $r_{1}$ from $RG_1$\;
 $\Delta\gets -\log(r_{1})/\Sigma$\;
  \For{$\mathbf{u}\in\set E$}{
 $p_{\mathbf{u}} \gets g_{\mathbf{u}}\left(\theta,W^\theta(s)\right)/\Sigma$\;
 }
 Divide the interval $(0,1)$ into $|\set E|$ sub-intervals of length
 $p_{\mathbf{u}}$, $\mathbf{u}\in\set E$\;
 Take $r_{2}$ from $RG_2$ and let $\overline{\mathbf{u}}$
 such that $r_{2}$ lies within the sub-interval of length $p_{\overline{\mathbf{u}}}$\;
  $W^\theta(s+\Delta)\gets W^\theta(s)+\overline{\mathbf{u}}$\;
  $s\gets s+\Delta$\;
 }
 \caption{Gillespie Direct Method}
 \label{algo_gil1}
\end{algorithm}

\medskip

\modif{Using Algorithm \ref{algo_gil1}, it possible to analyze the sensitivity of model output to ${\bf Z}$, the random vector controlling the intrinsic randomness. However, with this algorithm we cannot conduct a finer sensitivity analysis, by quantifying separately the impact of intrinsic randomness associated to each transition type. To fix this drawback, 
we introduce Algorithm \ref{algo_gil2} below, which is a slight modification of Gillespie First Reaction Method using pseudo-random number generators. Algorithm \ref{algo_gil2} uses as many random numbers as the number of transition types per step. It is thus possible from this algorithm to analyze separately the sensitivity of model output to intrinsic randomness associated to each transition type.}\\

\begin{algorithm}[H]
\SetAlgoLined
\SetKwInOut{Input}{Inputs}
  \SetKwInOut{Output}{Output}
  \Input{$t_{\mathrm{end}}$, $\theta$, $\; Z:=\{RG_\mathbf{u},\mathbf{u}\in \set{E}\}$}
  \KwData{$\xi_0$, $\set{E}$, $\{g_{\mathbf{u}},\mathbf{u}\in \set{E}\}$}
\Output{$\{W^{\theta}(s), s\in [0,t_{\mathrm{end}}]\}$ }
 Initialization:
 $s\gets 0$, $W^{\theta}(s) \gets \xi_0$\;
 \While{$s < t_{\mathrm{end}}$}{
  \For{$\mathbf{u}\in \set{E}$}{
 Take $r_{\mathbf{u}}$ from $RG_\mathbf{u}$\;
$a_\mathbf{u} \gets g_{\mathbf{u}}\left(\theta,W^{\theta}(s)\right)$\;
$\Delta_\mathbf{u} \gets \frac{-\log(r_\mathbf{u})}{a_\mathbf{u}}$
 }
 $\overline{\mathbf{u}} \gets \text{argmin}_\mathbf{u}\Delta_\mathbf{u}$\;
 $\Delta \gets \Delta_{\overline{\mathbf{u}}}$\;
  $W(s+\Delta)\gets W(s)+\overline{\mathbf{u}}$\;
  $s\gets s+\Delta$\;
 }
 \caption{Gillespie First Reaction Method}
 \label{algo_gil2}
\end{algorithm}

\medskip

\modif{As an alternative to Gillespie First Reaction Method, we review the simulation algorithm introduced in \cite{Kurtz2} (see also \cite{Kurtz1}), based on the so-called random-time change representation.} More precisely, the random state $W^{\theta}(t)$ can be expressed,
for every $t\geq 0$, through
\begin{linenomath}
\begin{equation*}
W^\theta(t)=W^\theta(0)+\sum_{\mathbf{u}\in\set{E}}Y_\mathbf{u}\left(
    \int_0^tg_\mathbf{u}(\theta,W^\theta(s))\mathrm{d}s\right)
  \mathbf{u},
  \label{eq:rtc}
\end{equation*}
\end{linenomath}
where $\{Y_{\mathbf{u}}(t),\,t\ge 0\}$, $\mathbf{u} \in \set{E}$, are independent unit-rate Poisson processes associated with transition  types $\mathbf{u}\in\mathbf{E}$.
\modif{The next reaction method generates exact sample paths while only needing one random number per step.} Introduced in \cite{Navarro} as a tool for doing complete GSA of chemical reaction network models \citep{lemaitre}, Algorithm \ref{algo_kurtz} below 
 is a slight modification of the Modified Next Reaction Method proposed by \cite{anderson}.\\ 

\begin{algorithm}[H]
\SetAlgoLined
\SetKwInOut{Input}{Inputs}
  \SetKwInOut{Output}{Output}
  \Input{$t_{\mathrm{end}}$, $\theta$, $\; Z:=\{RG_\mathbf{u},\mathbf{u}\in \set{E}\}$}
  \KwData{$\xi_0$, $\set{E}$, $\{g_{\mathbf{u}},\mathbf{u}\in \set{E}\}$}
\Output{$\{W^\theta(s), s\in [0,t_{\mathrm{end}}]\}$ }
 Initialization:
 
 \For{$\mathbf{u}\in \set{E}$}{
 Take $r_\mathbf{u}$ from $RG_\mathbf{u}$\;
 $t_\mathbf{u} \gets 0$, $t_\mathbf{u}^+ \gets -\log(r_\mathbf{u})$\;
 }
 $s\gets 0$, $W^{\theta}(s) \gets \xi_0$\;
 \While{$s < t_{\mathrm{end}}$}{
   \For{$\mathbf{u}\in \set{E}$}{
$a_\mathbf{u} \gets g_{\mathbf{u}}\left(\theta,W(s)\right)$; $\Delta_\mathbf{u} \gets \frac{t_\mathbf{u}^+-t_\mathbf{u}}{a_\mathbf{u}}$
 }
 $\overline{\mathbf{u}} \gets \text{argmin}_\mathbf{u}\Delta_\mathbf{u}$\;
 $\Delta \gets \Delta_{\overline{\mathbf{u}}}$\;
 $W(s+\Delta)\gets W(s)+\overline{\mathbf{u}}$\;
  $s\gets s+\Delta$\;
    \For{$\mathbf{u}\in \set{E}$}{
$t_\mathbf{u} \gets t_\mathbf{u}+a_\mathbf{u} \Delta$
 }
 Take $r_{\overline{\mathbf{u}}}$ from $RG_{\overline{\mathbf{u}}}$;
 
 $t_{\overline{\mathbf{u}}}^+ \gets t_{\overline{\mathbf{u}}}^+-\log(r_{\overline{\mathbf{u}}})$;
 }
 \caption{Modified Next Reaction Method}
 \label{algo_kurtz}
\end{algorithm}

\medskip

We introduced in this section three deterministic representations of compartmental models, each one based on a slight modification of an exact simulation algorithm: Gillespie Direct Method (Algorithm~\ref{algo_gil1}), Gillespie First Reaction Method (Algorithm~\ref{algo_gil2}) and Modified Next Reaction Method (Algorithm~\ref{algo_kurtz}). 
Now, the aim of Section \ref{covid} is to implement a complete GSA using each of the above representations on a SARS-CoV-2 spread model. \modif{Depending on the number of compartments in our model, the computational cost may vary from one algorithm to the other. However for this specific SARS-CoV-2 model, simulation times were comparable.} From the theoretical results of Section \ref{SAfree}, we expect GSA results to differ from one representation to the other.


\section{Application to a SARS-CoV-2 spread model}\label{covid}

\modif{We propose in this section, as a case study, to apply the methodology of global sensitivity analysis reviewed in the previous sections to a parsimonious SARS-CoV-2 spread model}, which is a
simplified but still realistic version of the model introduced
in \cite{cazelles}. We do not pretend to provide the
most suitable model for the propagation of SARS-CoV-2, we rather aim
at demonstrating the effectiveness of the approach presented in
Section \ref{sto_framework} for a complete GSA of stochastic
compartmental models. In order to illustrate the statement in Section
\ref{SAfree} that sensitivity analysis results depend on the
deterministic representation chosen for its implementation, we compare
the results by using each of Modified Gillespie Direct Method
(Algorithm \ref{algo_gil1}), Modified Gillespie First Reaction Method
(Algorithm \ref{algo_gil2}) and Modified Next Reaction Method
(Algorithm \ref{algo_kurtz}) for simulations. Recall that these
algorithms have been \modif{reviewed} in Section \ref{secrepr}. In Section
\ref{description} we describe the considered SARS-CoV-2 model. Then in
Section \ref{num_set} we introduce the quantities of interest and
detail our numerical setting for sensitivity analysis. Finally in
Section \ref{SA} we present the results of the sensitivity analyses
obtained from the different simulation algorithms of
Section~\ref{secrepr}. \modif{The code developed to perform the numerical
  experiments is available at
  \texttt{https://hal.inrae.fr/MATHNUM/hal-03565729}.}

\subsection{A SARS-CoV-2 spread model}
\label{description}
\begin{figure}[H]
\resizebox{14cm}{10cm}{
\begin{tikzpicture}[node distance=3cm,auto,>=latex']
    \node[int] (c) [] {$\mathbf{S}$};

    \node [int] (d) [right of=c, node distance=5cm] {$\mathbf{E}$};
    
 		 \node [int] (s) [draw=none, right of=d, node distance=4.7cm] {};
  
    \node [int] (e) [below of=s, node distance=4.8cm] {$\mathbf{I}$};
    
    \node [int] (g) [above of=s, node distance=4.2cm] {$\mathbf{A}$};
    
    \node [int] (h) [right of=e, node distance=6cm] {$\mathbf{H}$};  
    \node [int] (f) [right of=g, node distance=6cm] {$\mathbf{R}$};
   
    \node [int] (i) [below of=h, node distance=4.2cm] {$\mathbf{D}$};   

    \draw[->, ultra thick, black] (c) edge node {$\frac{\beta}{N}(W_I+W_ A)W_S$} (d);
    \draw[->, ultra thick, black] (d) edge node {$\gamma_E(1-p_{(E,A)})W_E$} (e);
    \draw[->, ultra thick, black] (e) edge node {$\gamma_I(1-p_C)W_I$} (f);
    \draw[->, ultra thick, black] (d) edge node {$\gamma_E\cdot p_{(E,A)}W_E$} (g);
    \draw[->, ultra thick, black] (g) edge node {$\gamma_AW_A$} (f);
    \draw[->, ultra thick, black] (e) edge node {$\gamma_I p_C(1-p_{D|C})W_I$} (h);
     \draw[->, ultra thick, black] (e) edge node {$\gamma_I p_C p_{D|C}W_I$} (i);
    \draw[->, ultra thick, black] (h) edge node [below, xshift = 1.5cm] {$ \quad \gamma_H(1-p_{(H,D)})W_H$} (f);
    \draw[->, ultra thick, black] (h) edge node {$\gamma_Hp_{(H,D)}W_H$} (i);
\end{tikzpicture}}
\centering
\caption{Compartmental model of the spread of SARS-CoV-2. The nodes are the possible health statuses and the arrows connecting them are the possible types of transition. The labels above the arrows are  the corresponding rate functions.}
\label{sars}.
\end{figure}
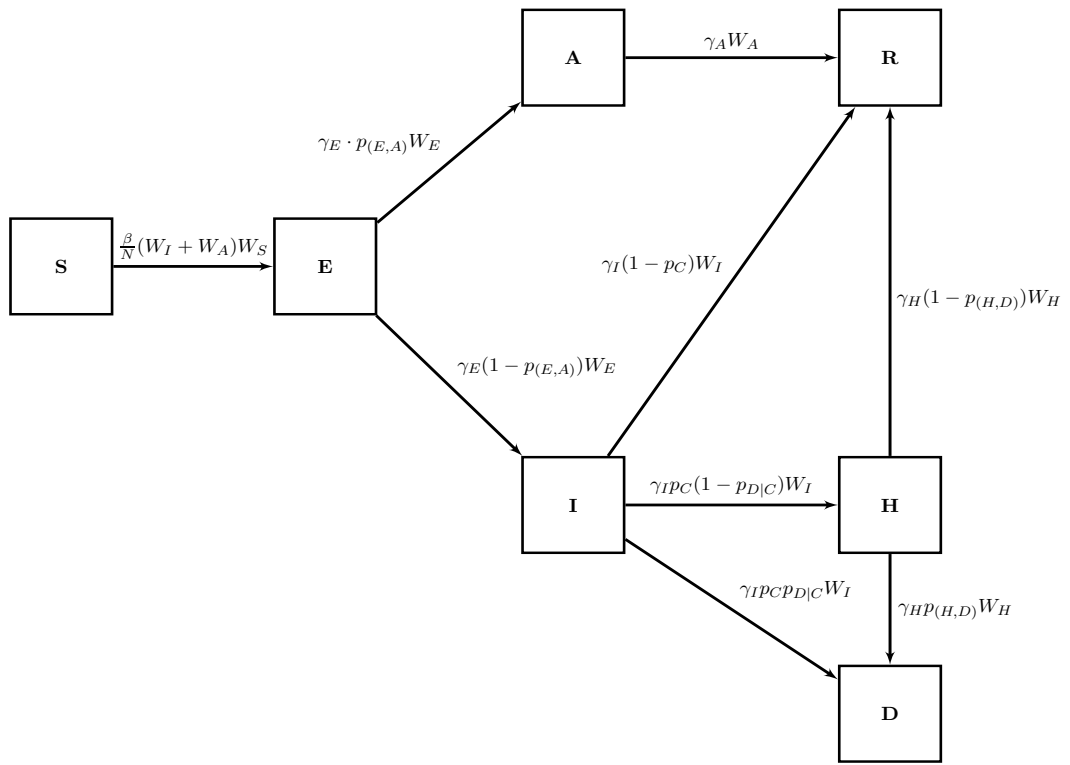

Recall from Section~\ref{sec:GsaForStoCompModels} that each process $W^{\theta}_{\alpha}$, $\alpha\in\mathbf{V}$, counts the number of individuals in compartment $\alpha$ over time. In this section we let $\mathbf{V}=\{S,E,A,I,H,R,D\}$, where the seven compartments represent seven possible health statuses: an individual can be susceptible (S), exposed (E) (i.e. infected but not yet infectious), asymptomatic infectious (A), symptomatic infectious (I), hospitalized (H), recovered (R) or dead (D). There are nine possible types of transition between these compartments,  see  
Figure~\ref{sars}. 
Note that infection is neglected within hospitals so that hospitalized individuals cannot infect.  Moreover, it is assumed that recovered individuals get perfectly immunized so they cannot be susceptible after recovering.
The vector of uncertain parameters is given by $\theta=\left(\beta,\gamma_E,\gamma_A,\gamma_{I},\gamma_H,p_{(E,A)}, p_C,p_{D|C},p_{(H,D)}\right)$. 
The different types of transition and their characteristics (transition vector $\mathbf{u}$ and associated rate function $g_{\mathbf{u}}$) are described in Table \ref{transitions}.

\modif{The list of uncertain epidemic parameters is given in Table \ref{parameters}. Instead of parameterizing the transitions from $I$ to $R$, $H$ or $D$ with the probabilities $p_{IR}$, $p_{IH}$ and $p_{ID}$, that correspond to the probability for an individual in $I$ to recover, to be hospitalized or to die, that are clearly not independent as $p_{IH}+p_{ID}+p_{IR}=1$, we chose to use instead parameters $p_C$ corresponding to the probability for an individual in compartment $I$ of being in a critical state and $p_{D|C}$ corresponding to the probability to die without being hospitalized conditionally on being in a critical state. These last two parameters are considered as independent from each other. This re-parameterization was inspired, e.g., from what is done in \cite[Chapter 7, page 191]{da2021basics}. More generally all the parameters listed in Table \ref{parameters} are assumed to be independent. The nominal value and range of variation for each parameter has been chosen in agreement with current knowledge and represent at least plausible values; see, e.g., \cite[Table S2 on page 15 of the Supplementary Material]{Knock2021} and, specifically for parameters $p_C$ and $p_{D|C}$, \cite[Chapter 7]{da2021basics}.}

\begin{table}[H]
\centering
\begin{tabular}{llll}
\hline
Transition   & Type & Transition vector $\mathbf{u}$ & Rate function $g_{\mathbf{u}}$ \\
\hline
 $ (S, E) $ & infection & $(-1,1,0,0,0,0,0)$ &$ \frac{\beta} {N} \cdot W_S\cdot \left(W_{A} + W_{I}\right) $ \\
 \hline
 $(E,A)$& \makecell[l]{asymptomatic\\ infectiousness activation} & $(0,-1,+1,0,0,0,0)$ & $\gamma_{E}\cdot p_{(E,A)}\cdot W_E$  \\
 \hline
 $(E,I)$ &  \makecell[l]{symptomatic \\ infectiousness activation} &  $(0,-1,0,+1,0,0,0)$ & $\gamma_{E}\cdot (1-p_{(E,A)})\cdot W_E$ \\
 \hline
$(A,R)$ & \makecell[l]{recovery of an \\asymptomatic }& $(0,0,-1,0,0,+1,0)$ &$\gamma_{A}\cdot W_{A}$ \\
\hline
$(I,R)$  & \makecell[l]{recovery of a \\symptomatic} & $(0,0,0,-1,0,+1,0)$ & $\gamma_{I}\cdot (1-p_C)\cdot  W_{I}$  \\
\hline
 $(I,H)$ & \makecell[l]{hospitalization \\of a symptomatic} & $(0,0,0,-1,+1,0,0)$ & $\gamma_{I}\cdot p_C \cdot (1-p_{D|C})\cdot W_{I}$ \\
\hline
 $(I,D)$ & \makecell[l]{death of a\\symptomatic} & $(0,0,0,-1,0,0,+1)$ & $\gamma_{I}\cdot p_C \cdot p_{D|C}\cdot W_{I}$ \\
\hline
$(H,R)$   &\makecell[l]{ recovery of a \\hospitalized} &   $(0,0,0,0,-1,+1,0)$ & $\gamma_H\cdot (1-p_{(H,D)})\cdot W_H$ \\
\hline
$(H,D)$  &   \makecell[l]{death of a \\hospitalized} & $(0,0,0,0,-1,0,+1)$ & $\gamma_H\cdot p_{(H,D)}\cdot W_H$  \\
\hline
\end{tabular}
\caption{Description of the model transitions between states $\{S,E,A,I,H,R,D\}$.}
\label{transitions}
\end{table}
\begin{table}[H]
\centering
\begin{tabular}{llll}
\hline 
Parameter & Description & Nominal value & Range of variation  \\ 
\hline 
$\beta$ & transmission rate  & $2.175$ & $(0.35,4)$\\ 
\hline 
$1/\gamma_E$ & mean sojourn duration in $E$ & $4.5$ days & $(2,7)$ \\ 
\hline 
$1/\gamma_{A}$ & mean sojourn duration in $A$ & 2 days & $(1,3)$\\ 
\hline 
$1/\gamma_{I}$  & mean sojourn duration in $I$  & $4$  days& $(3,5)$ \\ 
\hline 
$1/\gamma_H$ & mean sojourn duration in $H$ & 9.5 days &$(7,12)$ \\ 
\hline 
$p_{(E,A)}$ & \makecell[l]{probability for an exposed to \\become asymptomatic}  & $0.5$&$(0.3,0.7)$ \\ 
\hline 
$p_C$ & \makecell[l]{probability for an individual in\\ compartment $I$ of being in a critical state} & 0.175 & $(0.15,0.2)$ \\ 
\hline 
$p_{D|C}$ & \makecell[l]{probability to die without being hospitalized\\ knowing that the individual is in a critical state}  & 0.175  & $(0.15,0.2)$ \\ 
\hline
$p_{(H,D)}$ & \makecell[l]{probability for a hospitalized  to die}  & 0.0505 &$(0.001,0.1)$ \\ 
\hline 
\end{tabular} 
\caption{Model parameter nominal values and their range of variation in the sensitivity analysis.}
 \label{parameters}
\end{table}

\subsection{Setting for sensitivity analysis}
\label{num_set}
\begin{figure}[H]
\centering
\includegraphics[scale=0.20]{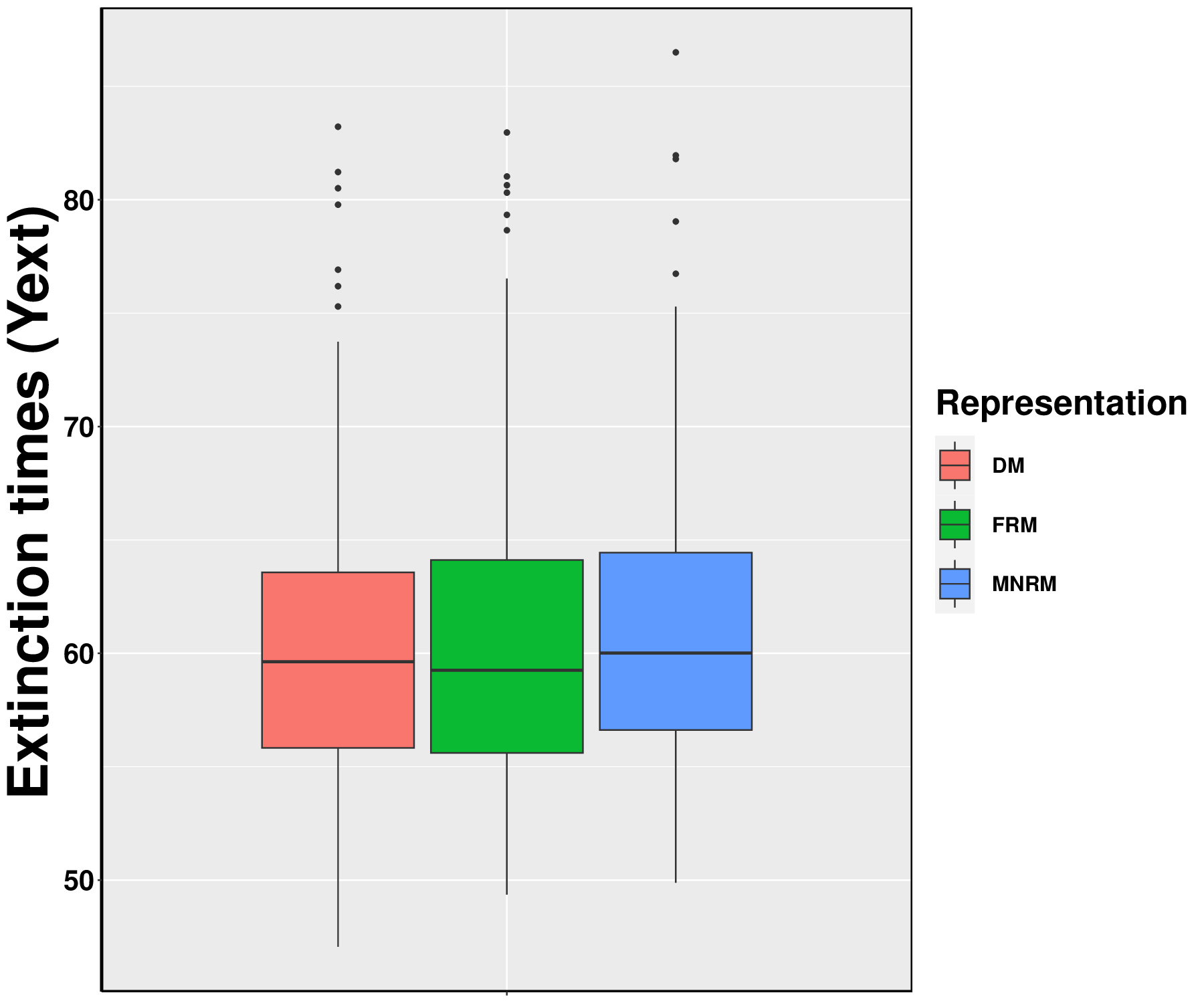}
\caption{Boxplot of $200$ simulations of $Y^{\theta}_{ext}$ performed with (left, red) Gillespie Direct Method (middle, green) Gillespie First Reaction (right, blue) Modified Next Reaction algorithm with uncertain parameters $\theta$ set to the nominal values given in Table~\ref{parameters}.}
\label{box_yext}
\end{figure}

We consider a population of $N=2005$ individuals including five exposed individuals at the start of the epidemic $t=0$, so that the process $W^\theta$ has the initial state
\begin{multline*}
\xi_0=\left(W_S(0)=2000,W_E(0)=5,\right.\\ \left. W_A(0)=W_I(0)=W_H(0)= W_R(0)= W_D(0)=0\right).
\end{multline*}
We focus on two quantities of interest (QoIs). First we consider a scalar QoI, namely the  extinction time $Y^{\theta}_{ext}$ of the epidemic,  defined as the first instant at which there are no exposed (E) nor infectious (A or I) individuals anymore:
  $$
 Y^{\theta}_{ext}=\inf\{t\geq 0: W^{\theta}_E(t)+W^{\theta}_A(t)+W^{\theta}_I(t)=0\}.
 $$
Note that for all $\theta\in \Theta$, $Y^{\theta}_{ext}$ is well-defined, i.e.  $Y^{\theta}_{ext}<+\infty$. Indeed, by considering the compartmental model described in Figure \ref{sars}, after a finite number of transitions, the stochastic process will necessarily reach an absorbing state with empty compartments $E,A$ and $I$.
We display in Figure~\ref{box_yext} two hundred independent realizations of $Y_{ext}^{\theta}$ for each of the simulation algorithms of Section~\ref{secrepr}. The uncertain parameters $\theta$ were set to the nominal values given in Table~\ref{parameters}. 
The three boxplots are  similar, which was expected since the distributions of the processes returned by each of the  three algorithms are the same. 

\begin{figure}[H]
     \centering
     \begin{subfigure}[b]{0.3\textwidth}
         \centering
         \caption{Direct Method}
         \includegraphics[width=\textwidth]{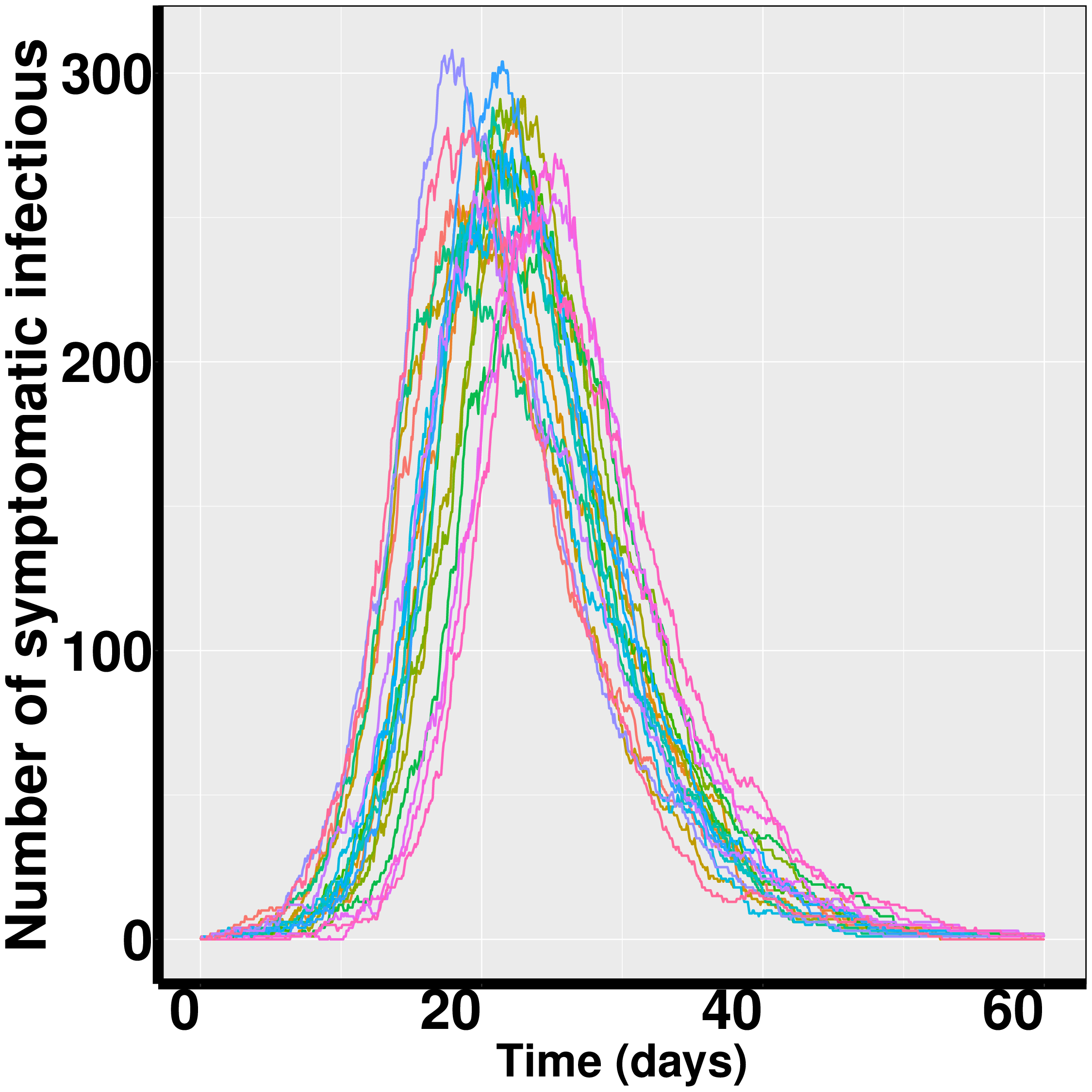} 
         \label{fig:Gil}
     \end{subfigure}
     \hfill
     \begin{subfigure}[b]{0.3\textwidth}
         \centering
         \caption{First Reaction Method}
         \includegraphics[width=\textwidth]{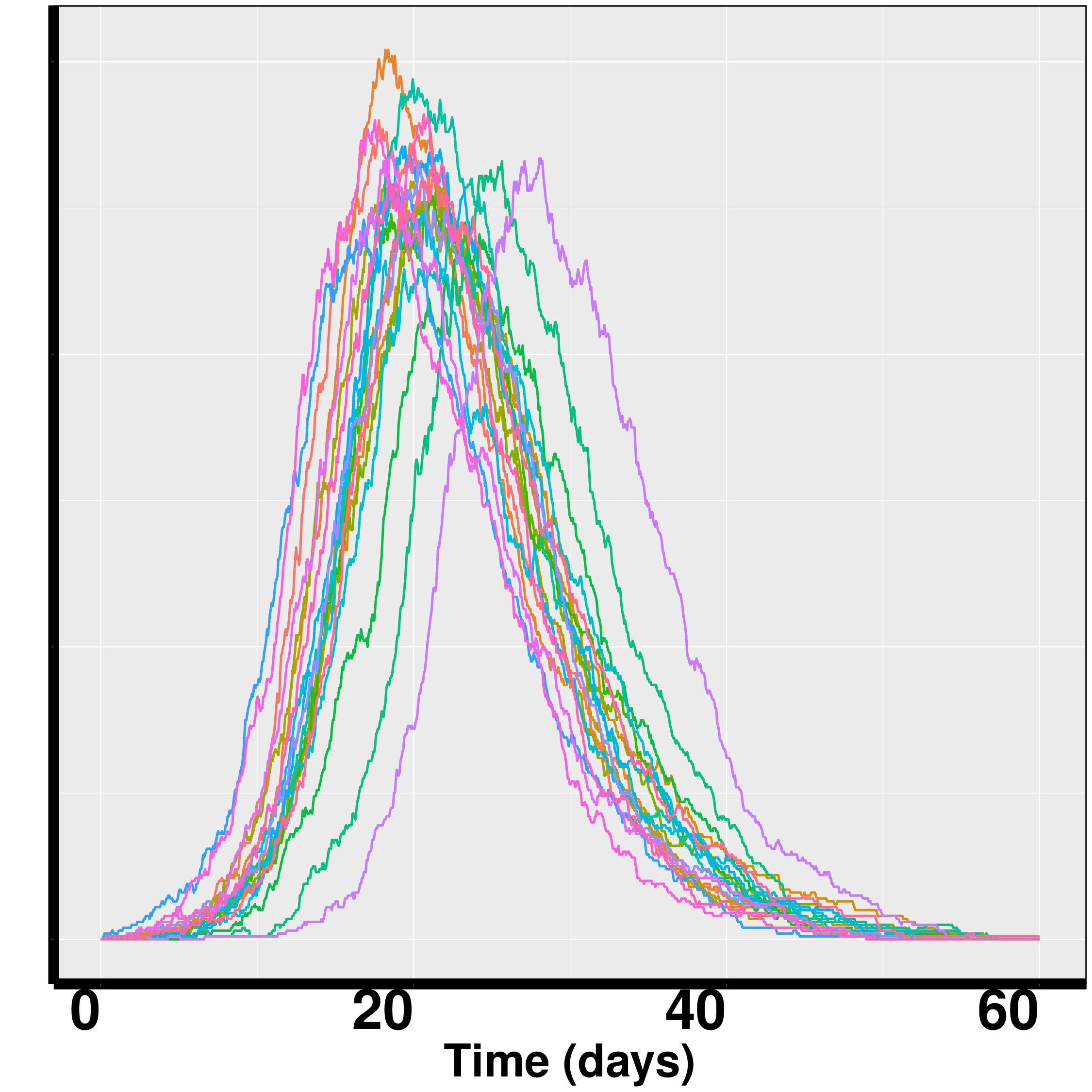}
         \label{fig:Gil2}
     \end{subfigure}
     \hfill
     \begin{subfigure}[b]{0.3\textwidth}
         \centering
         \caption{MNRM}
         \includegraphics[width=\textwidth]{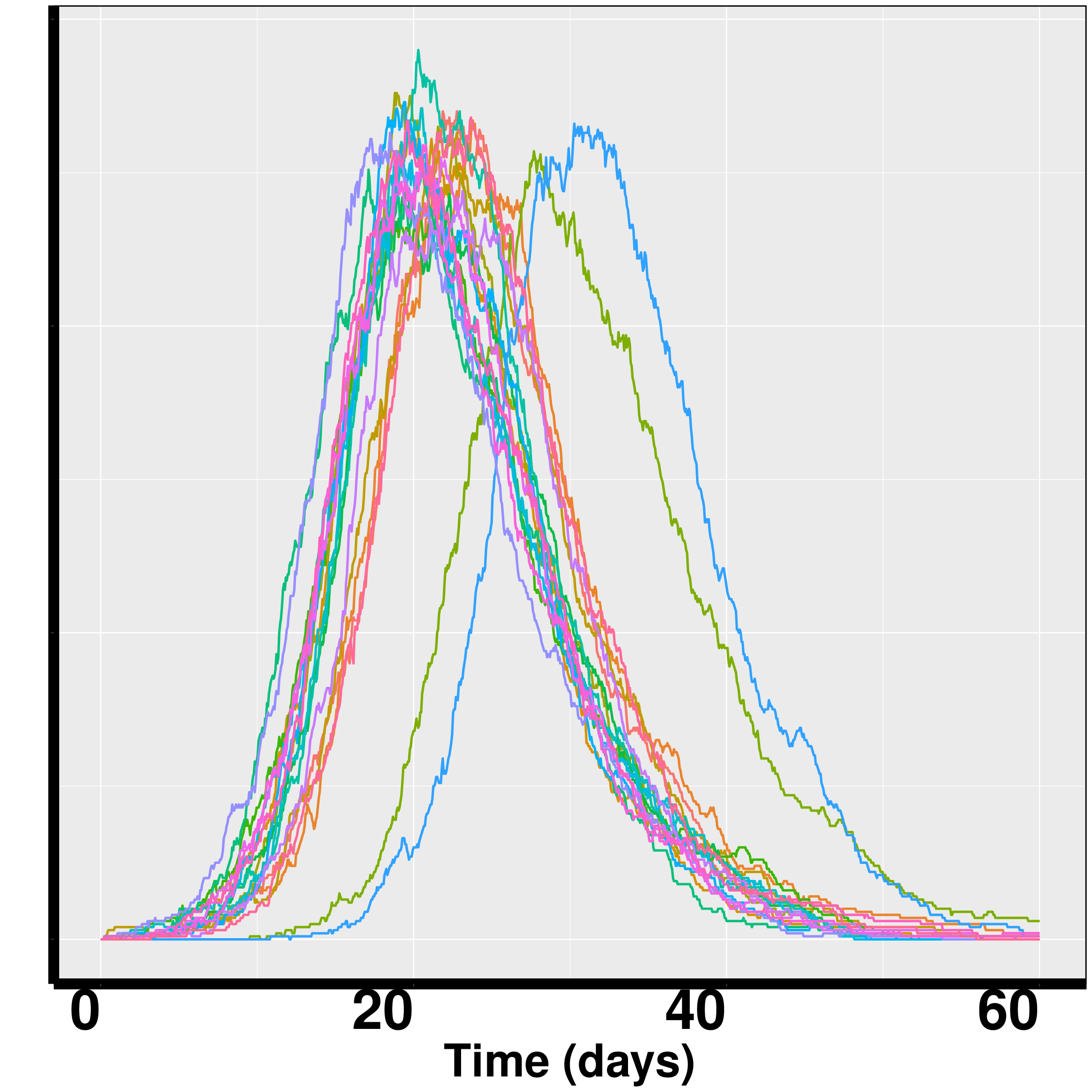}
         \label{fig:five over x}
     \end{subfigure}\\
     \begin{subfigure}[b]{0.32\textwidth}
         \centering
         \includegraphics[width=\textwidth]{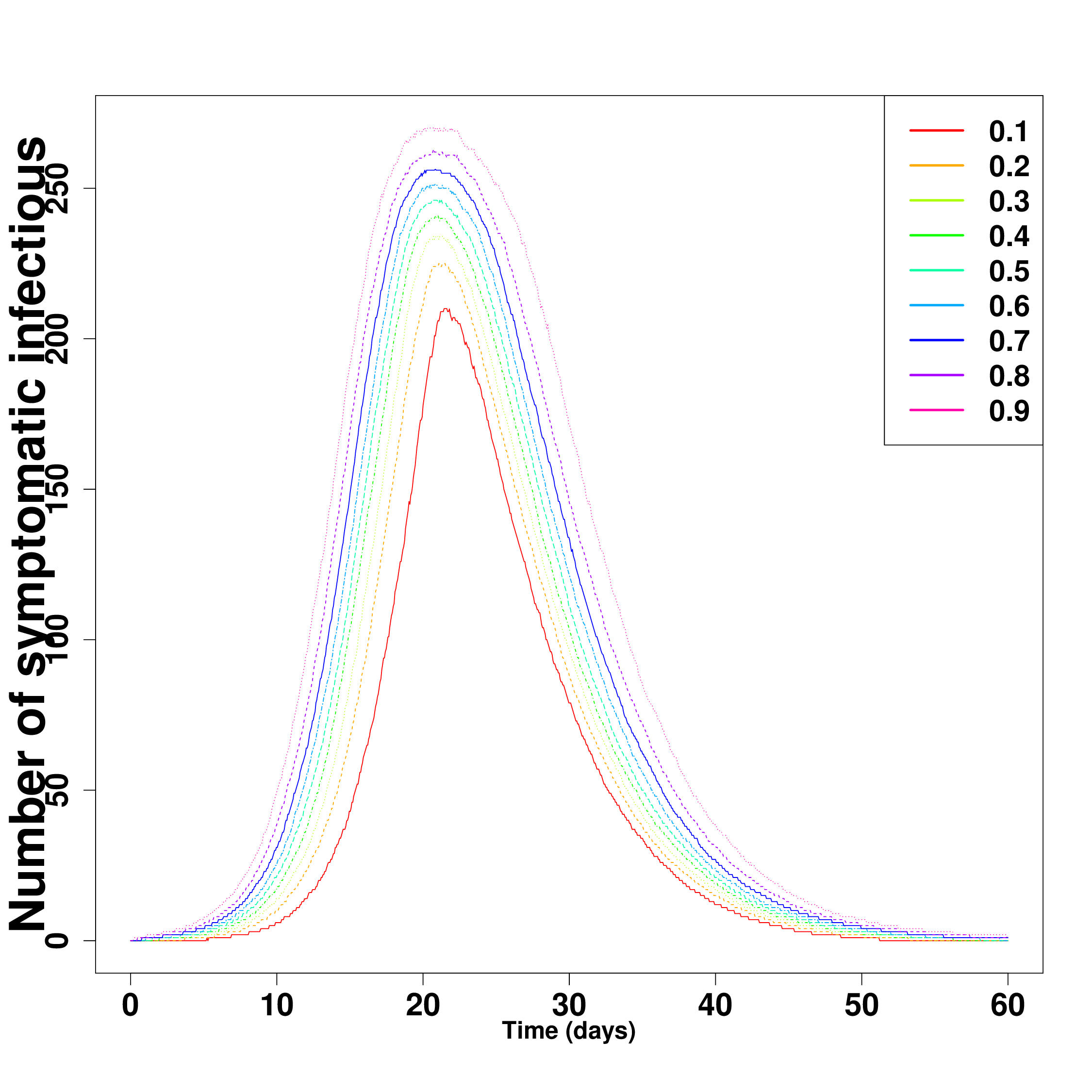}
         \label{fig:Gil_quant}
     \end{subfigure}
     \hfill
     \begin{subfigure}[b]{0.32\textwidth}
         \centering
         \includegraphics[width=\textwidth]{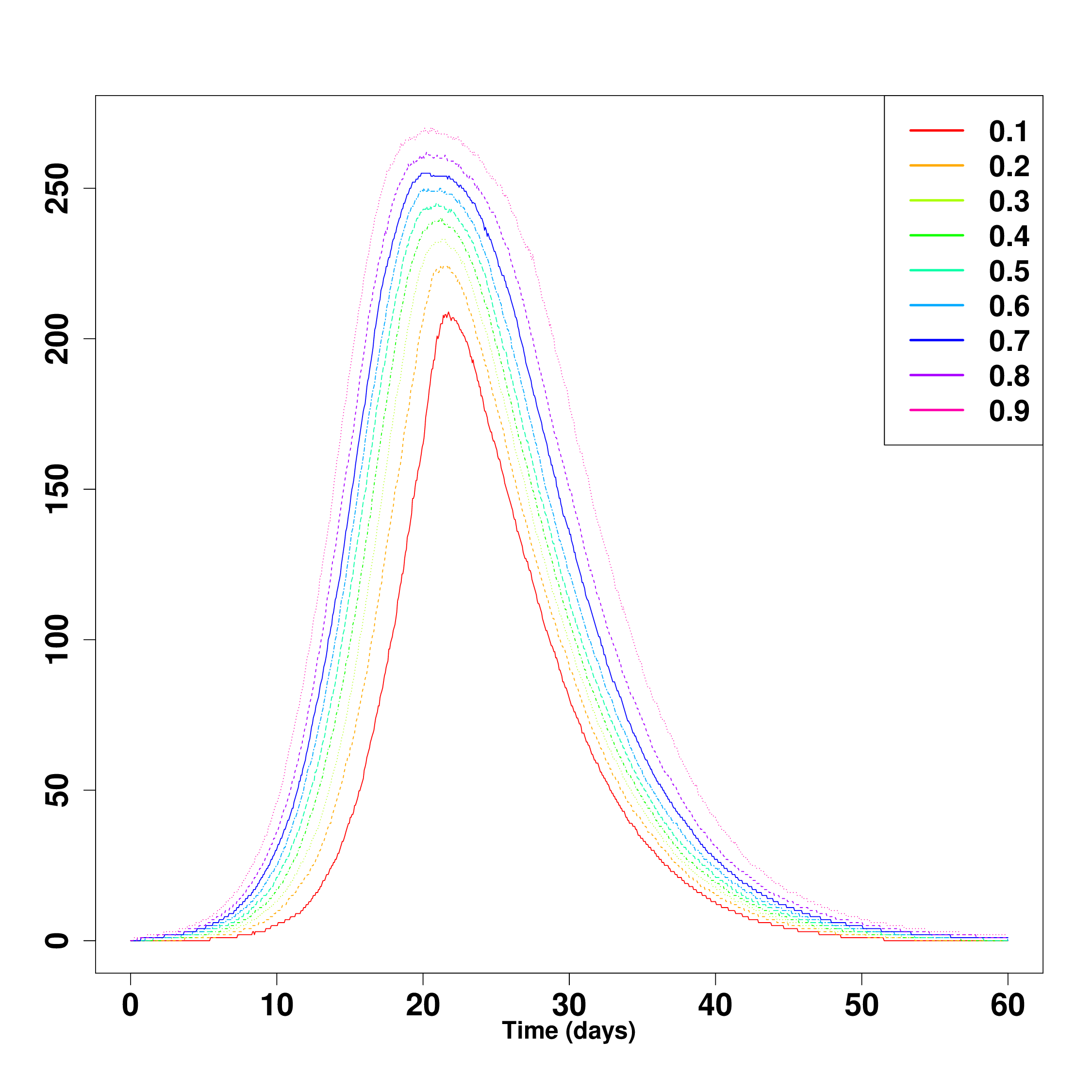}
         \label{fig:Gil2_quant}
     \end{subfigure}
     \hfill
     \begin{subfigure}[b]{0.32\textwidth}
         \centering
         \includegraphics[width=\textwidth]{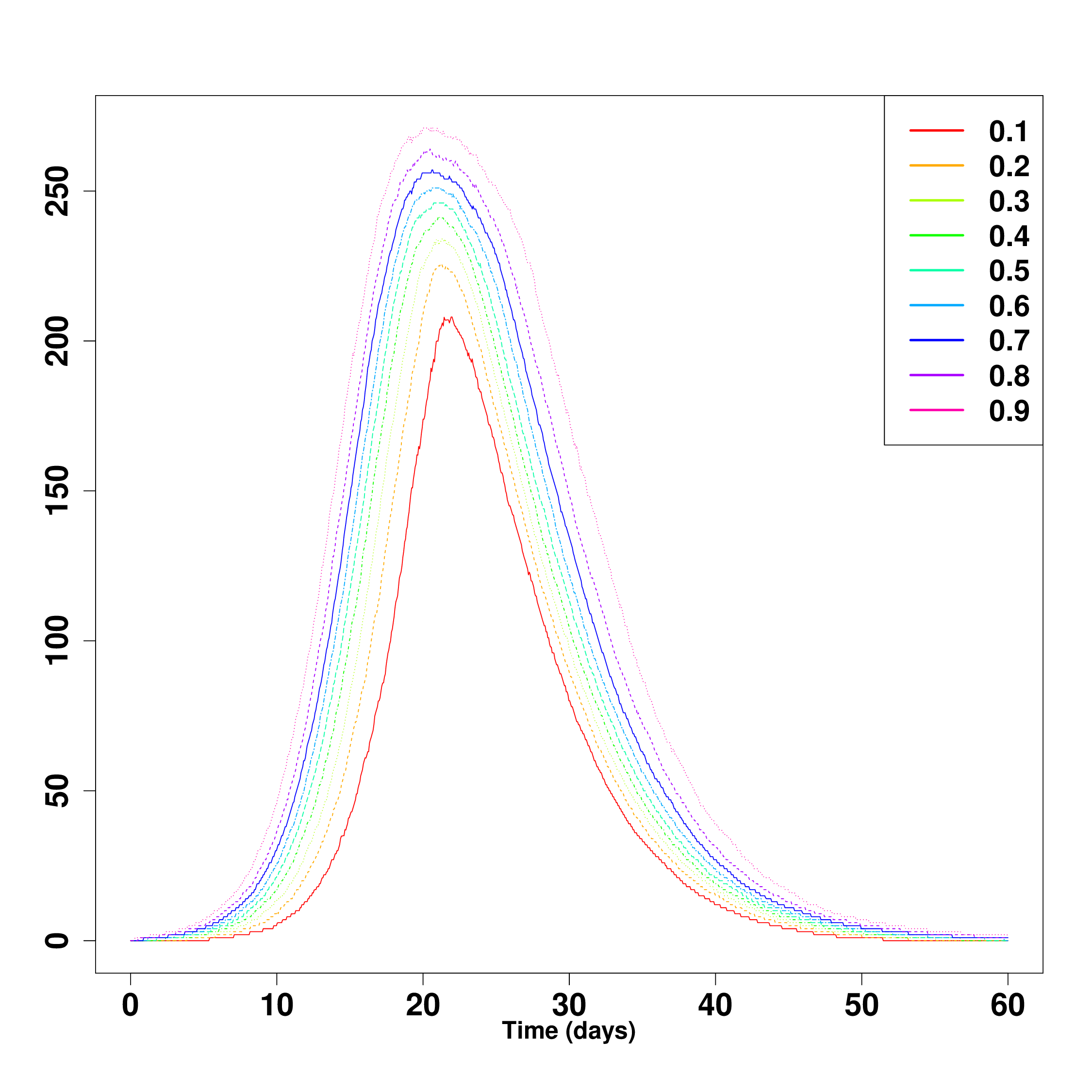}
         \label{fig:kurt_quant}
     \end{subfigure}
        \caption{\modif{First row: $20$ independent realizations of $t \rightarrow W^{\theta}_I(t)$. Second row: empirical quantile of level $0.1,\cdots  $ and $0.9$ as a function of time, computions are based on $1000$ independent sample paths of $t \rightarrow W^{\theta}_I(t)$. For each row, the components of the parameter vector $\theta$ were set to their nominal value (see Table \ref{parameters}) and simulations were obtained from Gillespie Direct Method (left), Gillespie First Reaction Method (middle), Modified Next Reaction Method (right).}}
        \label{YI}
\end{figure}

The second QoI we are considering is the dynamic of the number of symptomatic infectious individuals:
$$
Y^{\theta}_I=\{W^{\theta}_I(t), t\in[0,t_{\mathrm{end}}]\},
$$
where $t_{\mathrm{end}}$ was set to 60. (The process dies out at around that time, see Figure~\ref{YI}.) 
We display in Figure~\ref{YI} twenty independent  realizations of the process $Y_{I}^{\theta}$ for each of the simulation algorithms of Section~\ref{secrepr}. The input parameter vector $\theta$ was set to the nominal values given in Table~\ref{parameters}. 
The three charts in Figure~\ref{YI} display similar sample paths for $Y_I^\theta$, which was expected since the distribution of the processes returned by each of the  three algorithms is the same. 

\modif{In the top row of Figure \ref{YI}, we show $20$ independent trajectories of the process $t \mapsto W_I^{\theta}(t)$, obtained from Gillespie Direct Method (left), Gillespie First Reaction Method (middle) and Modified Next Reaction Method (right) with the components of the parameter vector $\theta$ fixed to their nominal value (see Table \ref{parameters}). On the bottom row of the same figure, we show the evolution over time of quantiles of different order (from $0.1$ to $0.9$) calculated from $1000$ independent trajectories of each algorithm (from left to right). These plots are in coherence with the fact that Algorithms \ref{algo_gil1}, \ref{algo_gil2} and \ref{algo_kurtz} are all exact simulation algorithms of the same stochastic process.}

In practice, simulations are carried out using the \RR Statistical Software \citep{Rsoft}. Sensitivity indices are estimated by using the \RR package \textsf{sensitivity} \citep{sensitivity}.  The function \textsf{soboljansen()} is used for total Sobol' index estimation while \textsf{sobol2007()} is used for first-order Sobol' index estimation. A priori distributions for model parameters are uniform distributions as described in Table \ref{parameters} and the a priori for intrinsic randomness is modeled by seeds uniformly distributed in $\{1,\cdots ,10^9\}$. Sensitivity indices are estimated from two independent designs of $n=2000$ input-output samples, where for each sample  a trajectory of $W^\theta$ is simulated through either Algorithm \ref{algo_gil1}, Algorithm \ref{algo_gil2} or Algorithm \ref{algo_kurtz}. As the dimension of the input space is large (at least $7$ model parameters plus inputs modeling intrinsic randomness whose number depends on the simulation algorithm), we use Latin Hypercube Sampling (see, e.g., \cite{lintan15}). Latin Hypercube Samples are generated by using the \RR package \textsf{DiceDesign} \citep{dicedesign}.

\subsection{Sensitivity analysis results}
\label{SA}

This section is devoted to the presentation and comparison of sensitivity analysis results obtained for the algorithms presented in Section \ref{secrepr}, namely Gillespie Direct Method, Gillespie First Reaction Method and Modified Next Reaction Method. In Section \ref{scalar_output} we present the results for the scalar output of interest, namely the extinction time of the epidemy $Y_{ext}^{\theta}$. Then in Section \ref{sec:dyn} we present the sensitivity analysis results for the functional output corresponding to the dynamic of the number of symptomatic infectious individuals $Y_I^{\theta}$. Finally in Section \ref{sec:choice} we discuss the choice of algorithms, depending on the practitioner's objectives.
Recall that in all the results presented in this section, sensitivity indices were estimated
from two independent designs of $n=2000$ input-output samples, and the estimation was repeated independently $50$ times for the different boxplots.

\subsubsection{Sensitivity analysis results for $Y_{ext}^{\theta}$}
\label{scalar_output}

\begin{figure}[H]
     \centering
     \begin{subfigure}[b]{1\textwidth}
         \centering
        \includegraphics[scale=0.25]{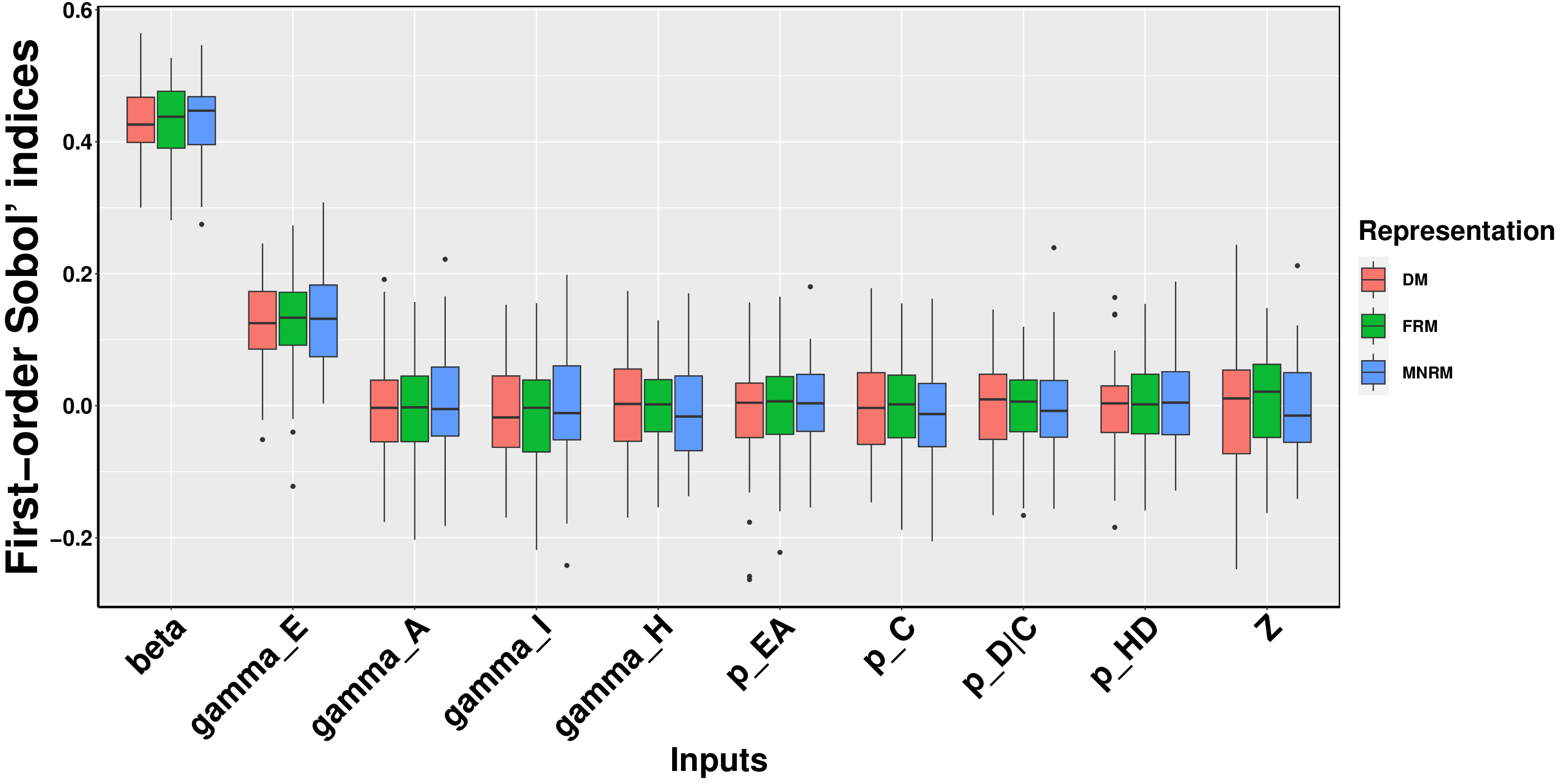} 
     \end{subfigure}
     \\
     \begin{subfigure}[b]{1\textwidth}
         \centering
         \includegraphics[scale=0.25]{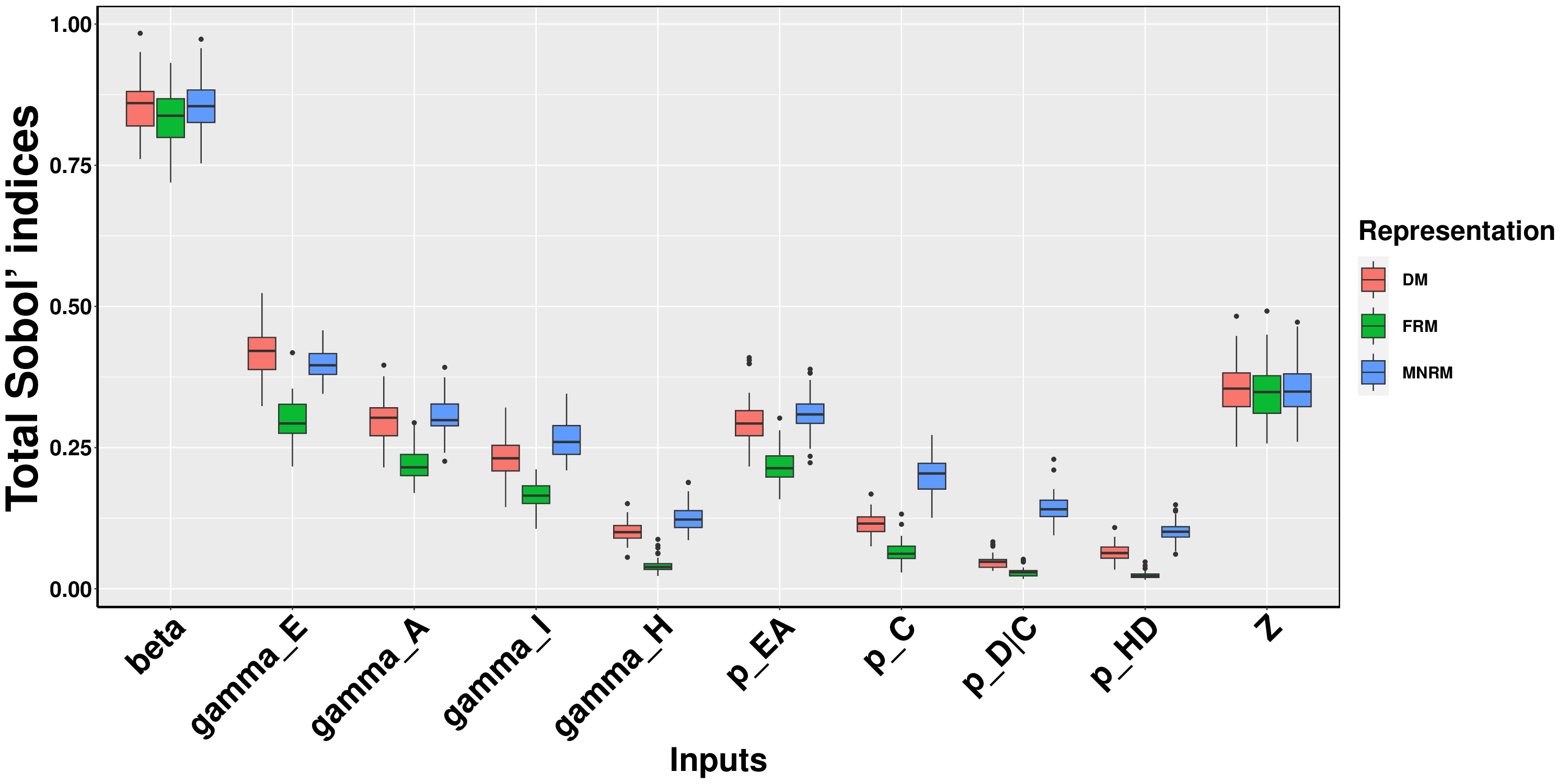}
     \end{subfigure}
         \caption{Boxplots of 50 independent estimates of the (top) first-order and (bottom)  total Sobol' indices for $Y_{ext}^\theta$, each computed with two independent designs of $n=2000$ input-output samples, for each simulation algorithm: (red, left) Gillespie Direct Method, (green, middle) Gillespie First Reaction, (blue, right) Modified Next Reaction.}
    \label{extinction}
\end{figure}

We display on Figure \ref{extinction} boxplots of first-order and total Sobol' index estimates. 

In accordance with the results stated in Section \ref{SAfree}, we observe on the top of Figure \ref{extinction} that there are no significant differences between the three algorithms for the first-order Sobol' index estimates. The only input parameters with a significant first-order effect are $\beta$ and $\gamma_E$. The sum of the first-order index estimates is far below $1$ which means that interactions are not negligible. 
We observe on the bottom of Figure \ref{extinction} that almost all inputs have a total effect significantly greater than zero. The interaction strength varies from one simulation algorithm to the other. This is due to the fact that the modeling of intrinsic randomness depends on each simulation algorithm. In particular, we observe that the total Sobol' index estimates corresponding to the Modified Next Reaction algorithm are never less than their counterpart computed from the First Reaction Method algorithm. 
This reflects a stronger interaction with intrinsic random noise for
Modified Next Reaction algorithm.
Finally, as expected from the theoretical results in Section~\ref{SAfree}, the total index estimates associated with intrinsic randomness do not depend on the chosen simulation algorithm.

\subsubsection{Sensitivity analysis results for $Y_{I}^\theta$}\label{sec:dyn}

\begin{figure}[H]
  \centering
  \begin{subfigure}[b]{1\textwidth}
    \centering
    \includegraphics[scale=0.3]{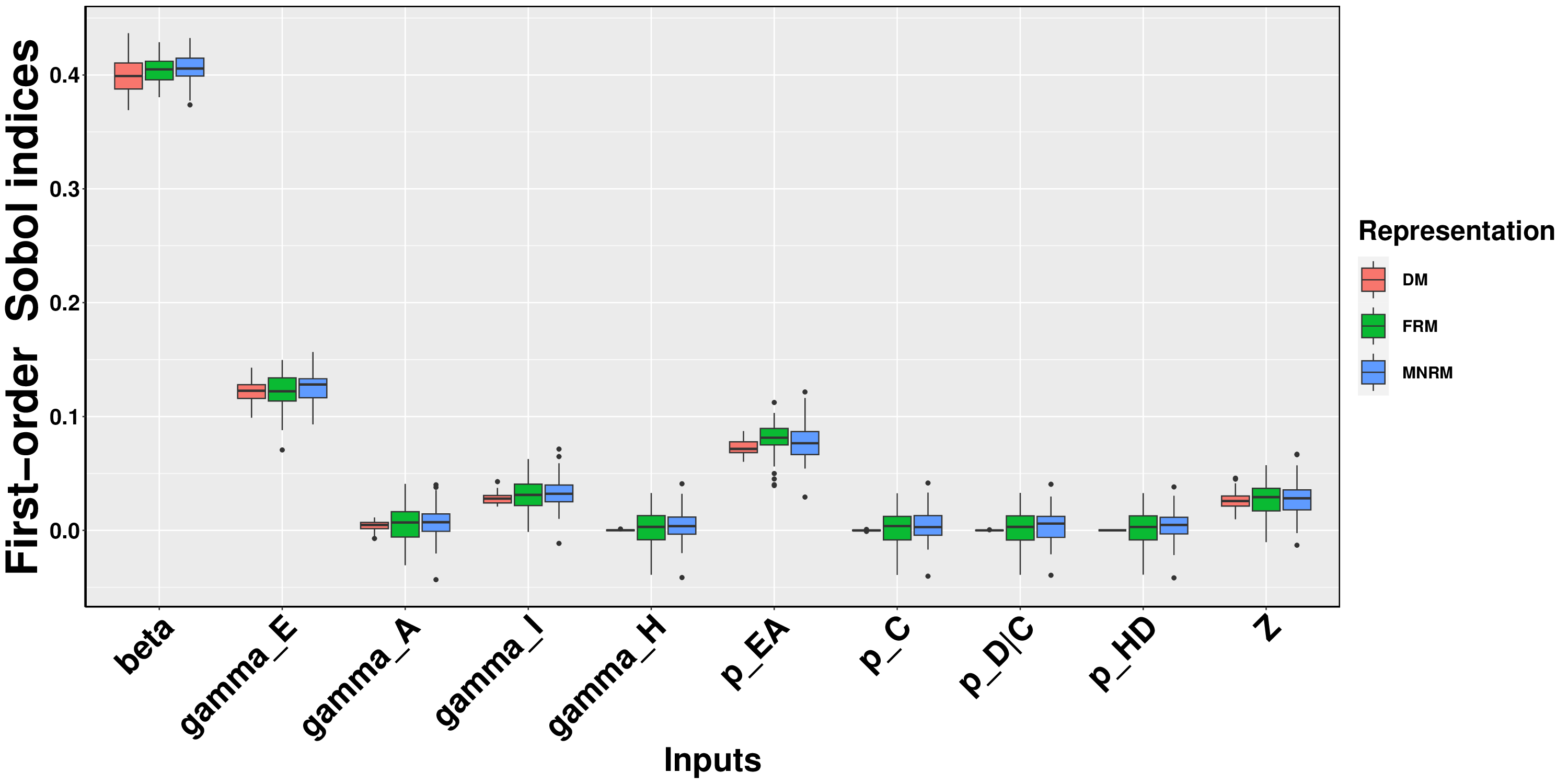} 
  \end{subfigure}
  \\
  \begin{subfigure}[b]{1\textwidth}
    \centering
    \includegraphics[scale=0.3]{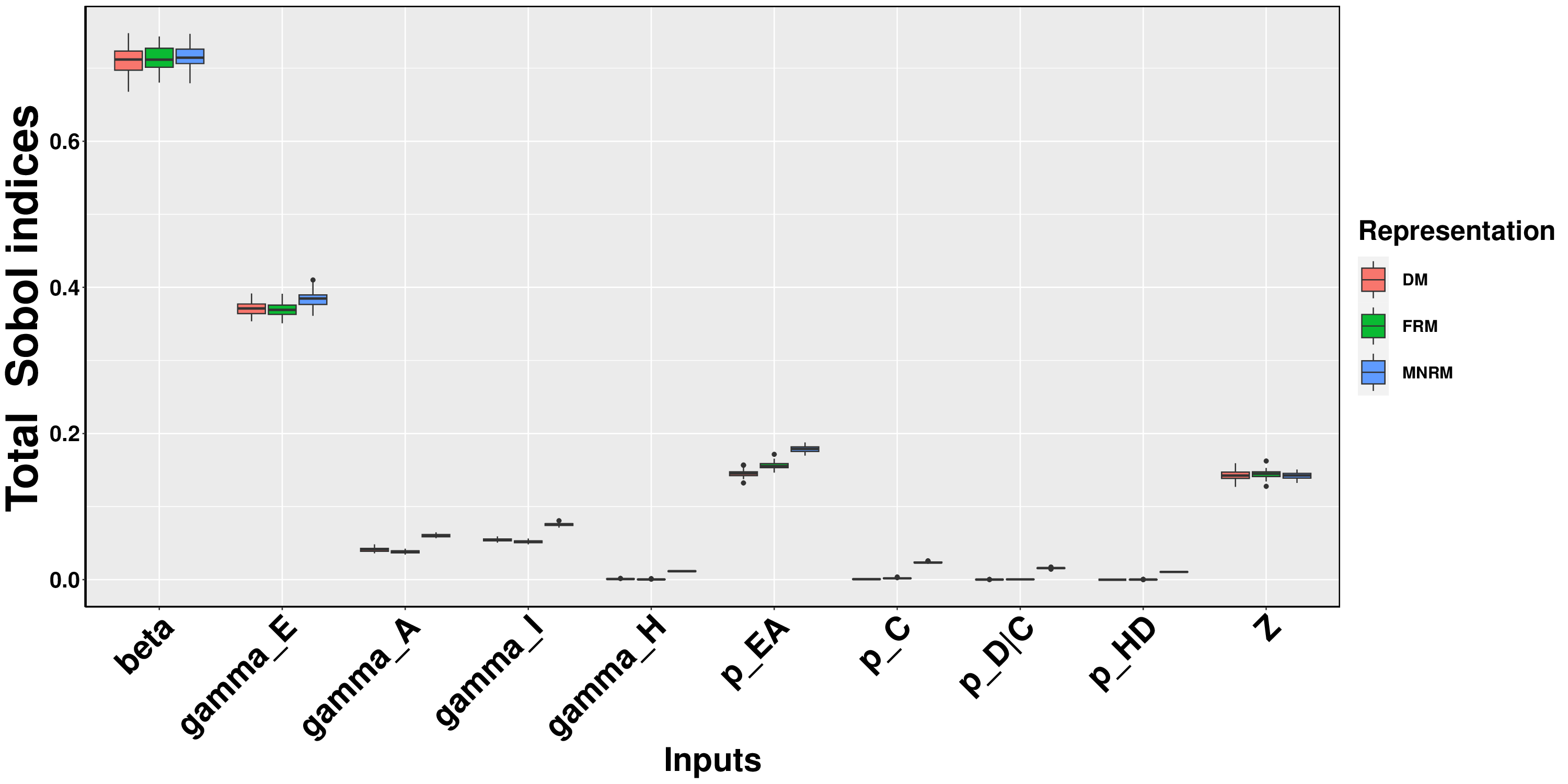}
  \end{subfigure}
  \caption{Aggregated (top) first-order and (bottom) total Sobol'
    index estimates for $Y_I^\theta$. For each
    input parameter (x-axis), boxplots are displayed for each
    simulation algorithm: (red,left) Gillespie Direct Method,
    (green,middle) Gillespie First Reaction, (blue,right) Modified
    Next Reaction. Each boxplot represents $50$
      independent index estimates, each of them computed with two independent designs of
      $n=2000$ input-output samples.}
  \label{agg}
\end{figure}

\begin{figure}[H]
     \centering
     \begin{subfigure}[b]{\textwidth}
         \centering
        \includegraphics[scale=0.25]{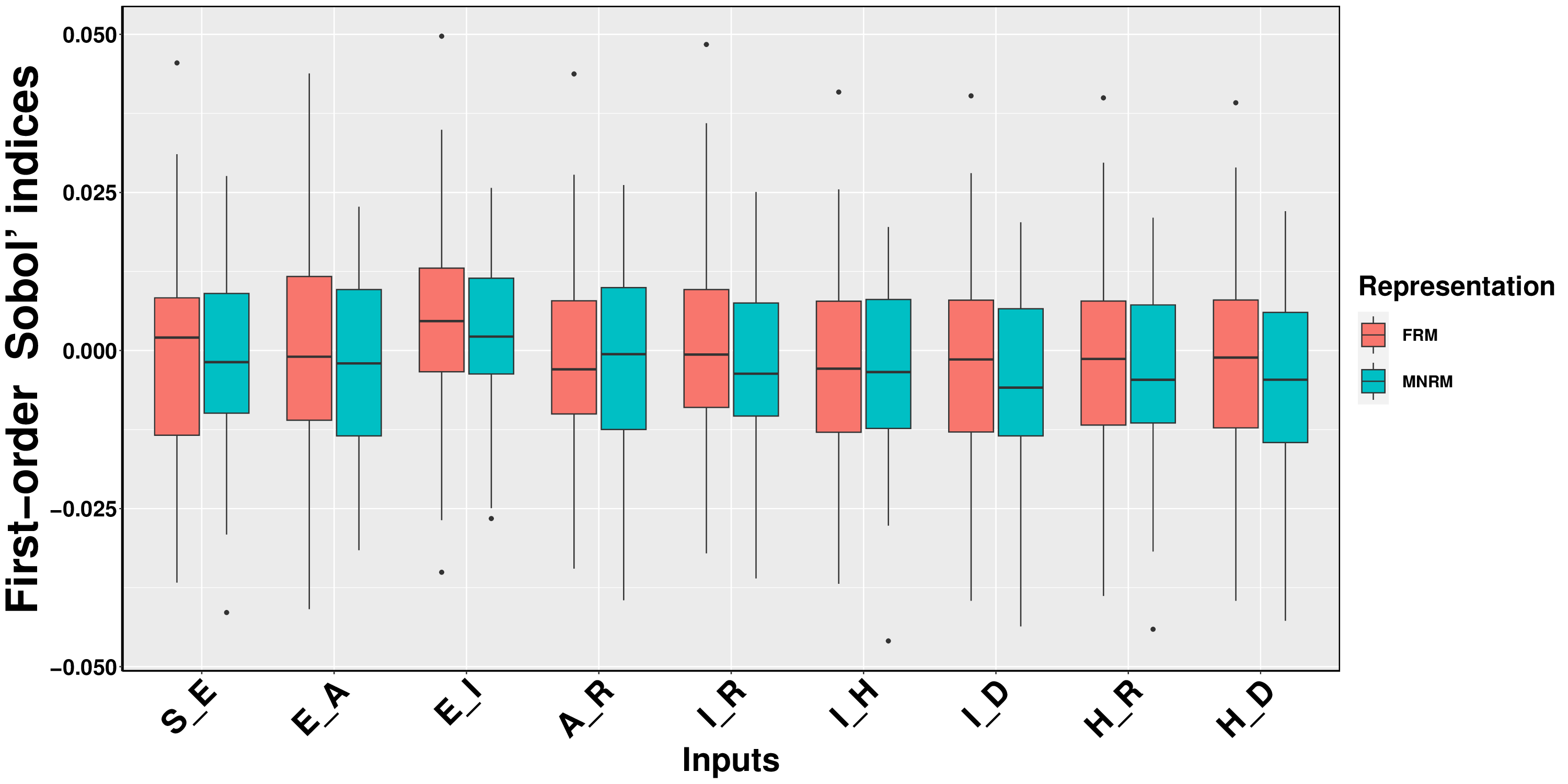} 
     \end{subfigure}
     \\
     \begin{subfigure}[b]{\textwidth}
         \centering
         \includegraphics[scale=0.25]{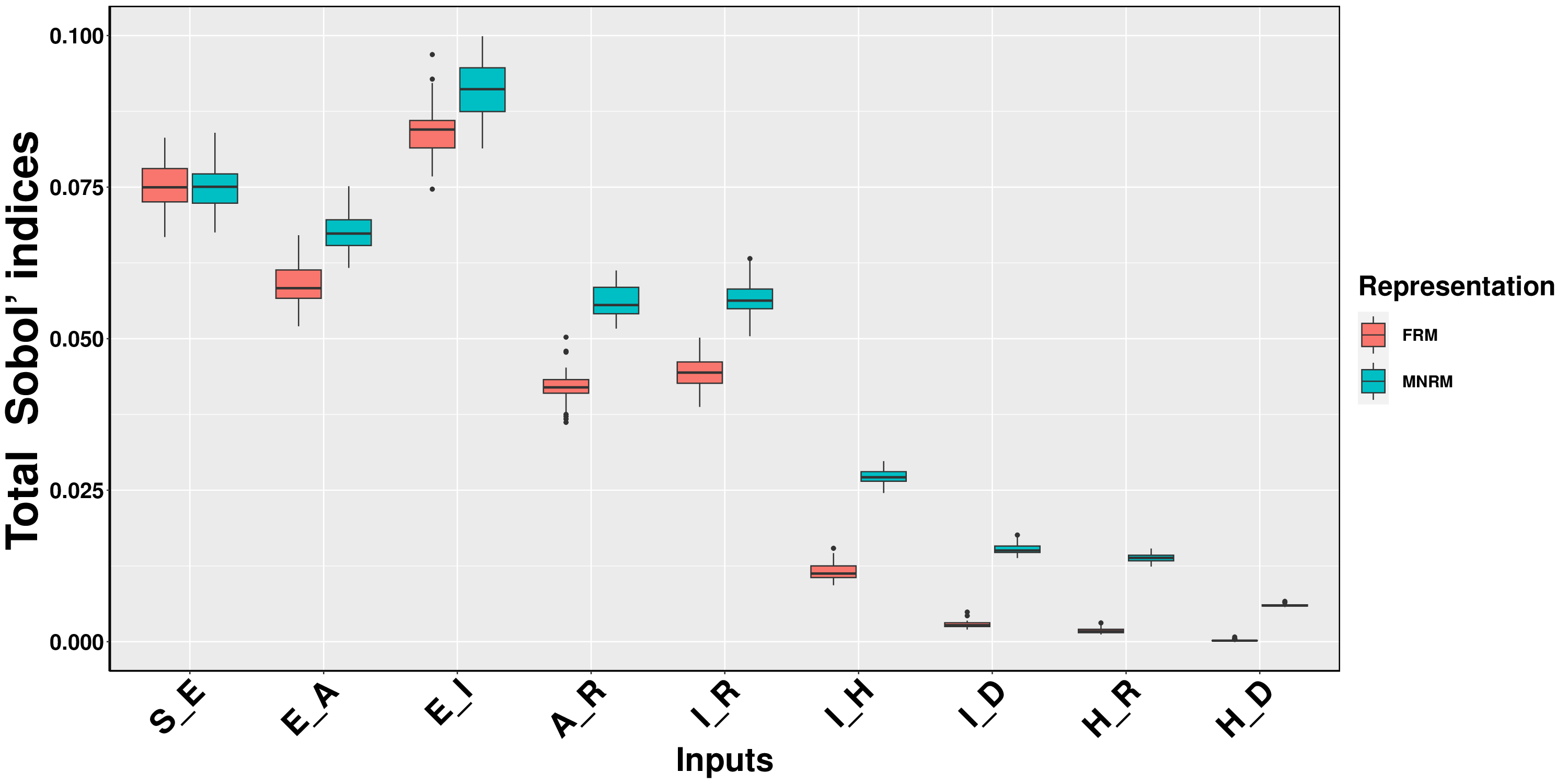}
     \end{subfigure}
     \caption{Aggregated (top) first-order and (bottom) total (b)
       Sobol' indices associated to each component of $Z$ (each type of transition described in Table
       \ref{transitions}). For each type of transition (x-axis),
       boxplots are displayed for (red,left) Gillespie First Reaction
       and (blue,right) Modified Next Reaction. Each boxplot
       represents $50$ independent index estimates, each of them
       computed with two independent designs of $n=2000$ input-output
       samples.}
    \label{aggtrans}
\end{figure}

Since $Y_I^\theta$ is a dynamical process, we can consider the
sensitivity of the whole trajectory or the sensitivity time by
time. In the numerical experiments, $Y_I^\theta$ is discretized over a
regular grid of size $1000$ of the interval
$[0,t_{\mathrm{end}}]$. The sensitivity of the whole trajectory
consists of computing estimates of the aggregated sensitivity indices
of Section~\ref{gsa}. These provide a scalar summary for the dynamical
evolution of first-order and total Sobol' indices. They are displayed
in Figure~\ref{agg}.  While the three algorithms show similar
first-order Sobol index estimates, they show some significant
differences for the total index estimates.  We observe that the total
index estimates for the uncertain parameters
$\gamma_A,\gamma_I,\gamma_H,p_{E,A},p_C,p_{D|C}$ and $p_{H,D}$ are
significantly higher for Modified Next Reaction Method, indicating
that each of those parameters interacts more with the variable $Z$.
Then, using Algorithms \ref{algo_gil2} or \ref{algo_kurtz}, it is
possible to decompose $Z$ into components that correspond to the
different types of transition.  On Figure \ref{aggtrans}, we plotted
first-order and total Sobol' index estimates associated with each of
those components, for both algorithms. While there were no difference
between the three algorithms for total index estimates associated with
the intrinsic noise $Z$ as a whole (see the bottom of Figure \ref{agg}), the analysis
by type of transition reveals that the total sensitivity estimates of its components
significantly differ from one algorithm to the other (see the bottom
of Figure \ref{aggtrans}).

\begin{figure}[H]
\setlength{\tempwidth}{.3\linewidth}
\settoheight{\tempheight}{\includegraphics[scale=0.1]{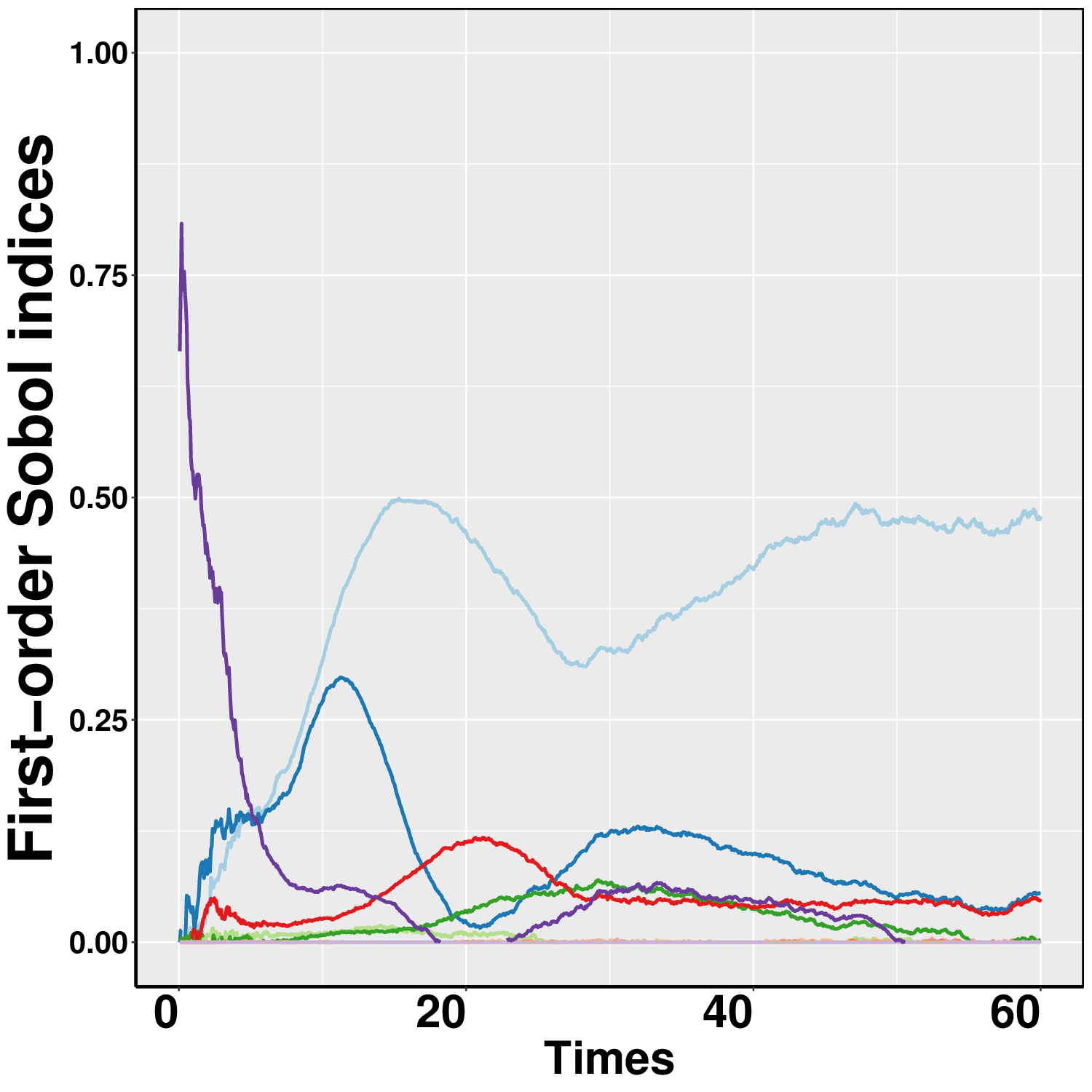}}%
     \centering
     \begin{tabular}{@{}c@{ }c@{ }}
        (a) & (b) \\
    
        \includegraphics[scale=0.22]{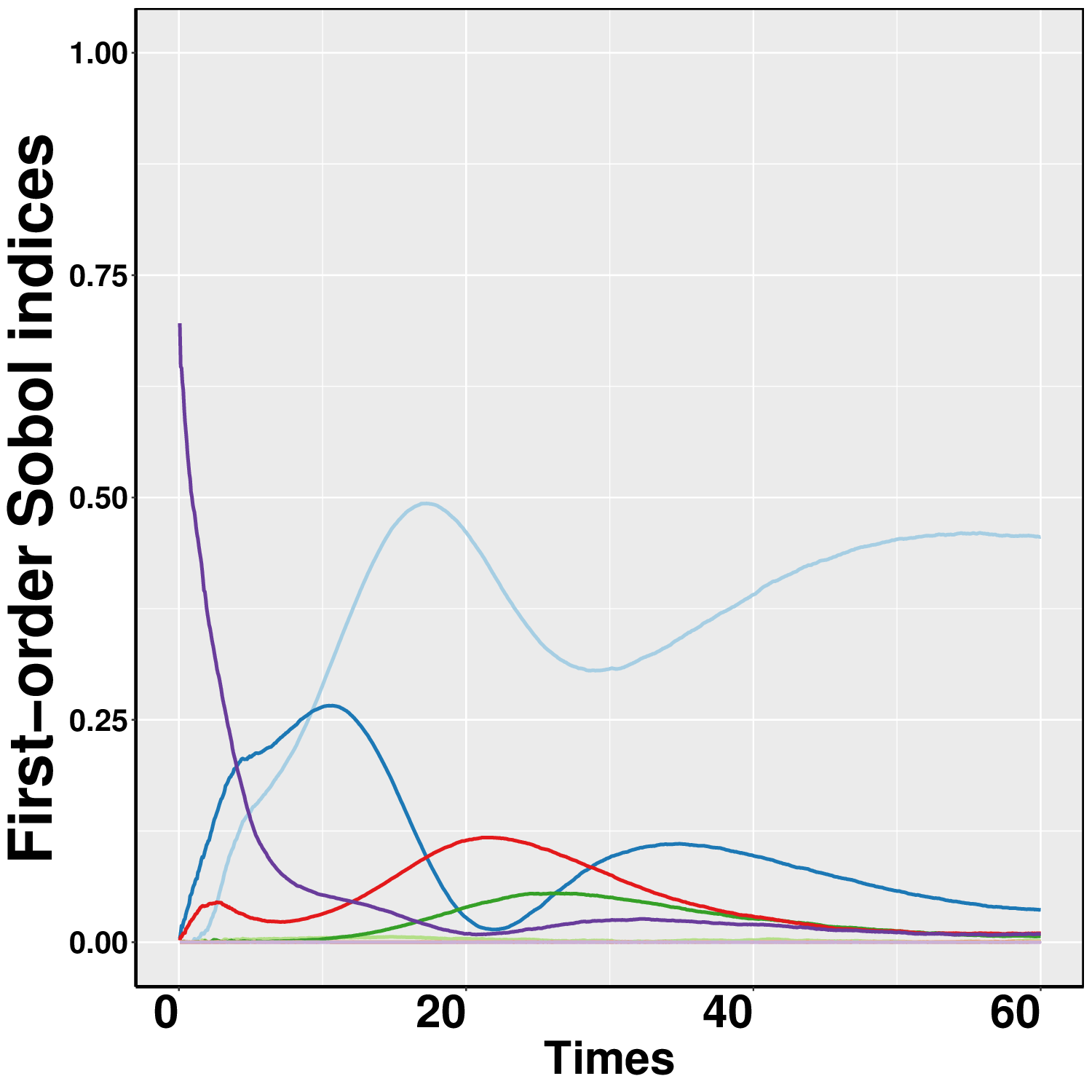} 
   &  
         \includegraphics[scale=0.22]{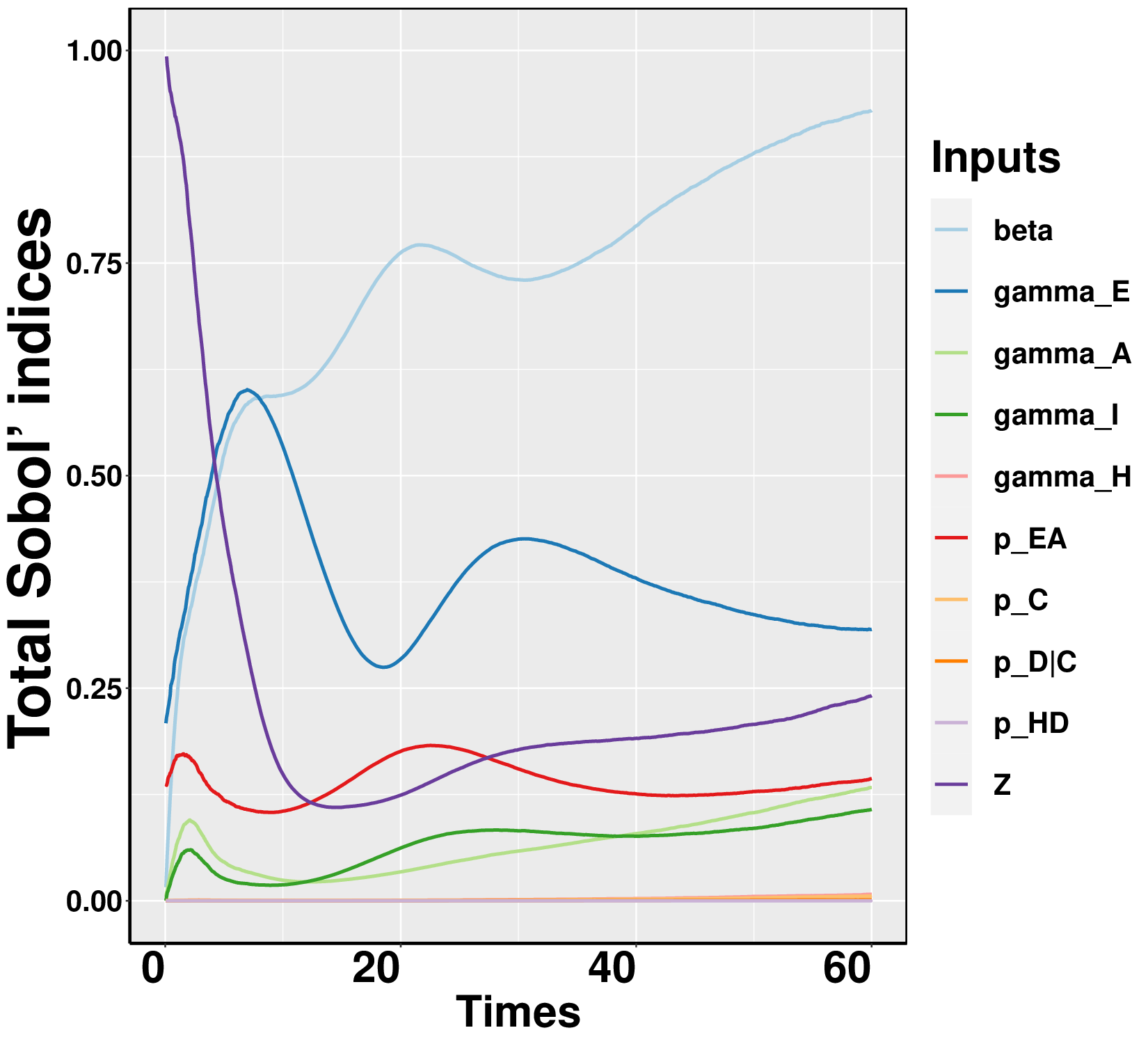}
           \\
            (c) & (d) \\
            \includegraphics[scale=0.22]{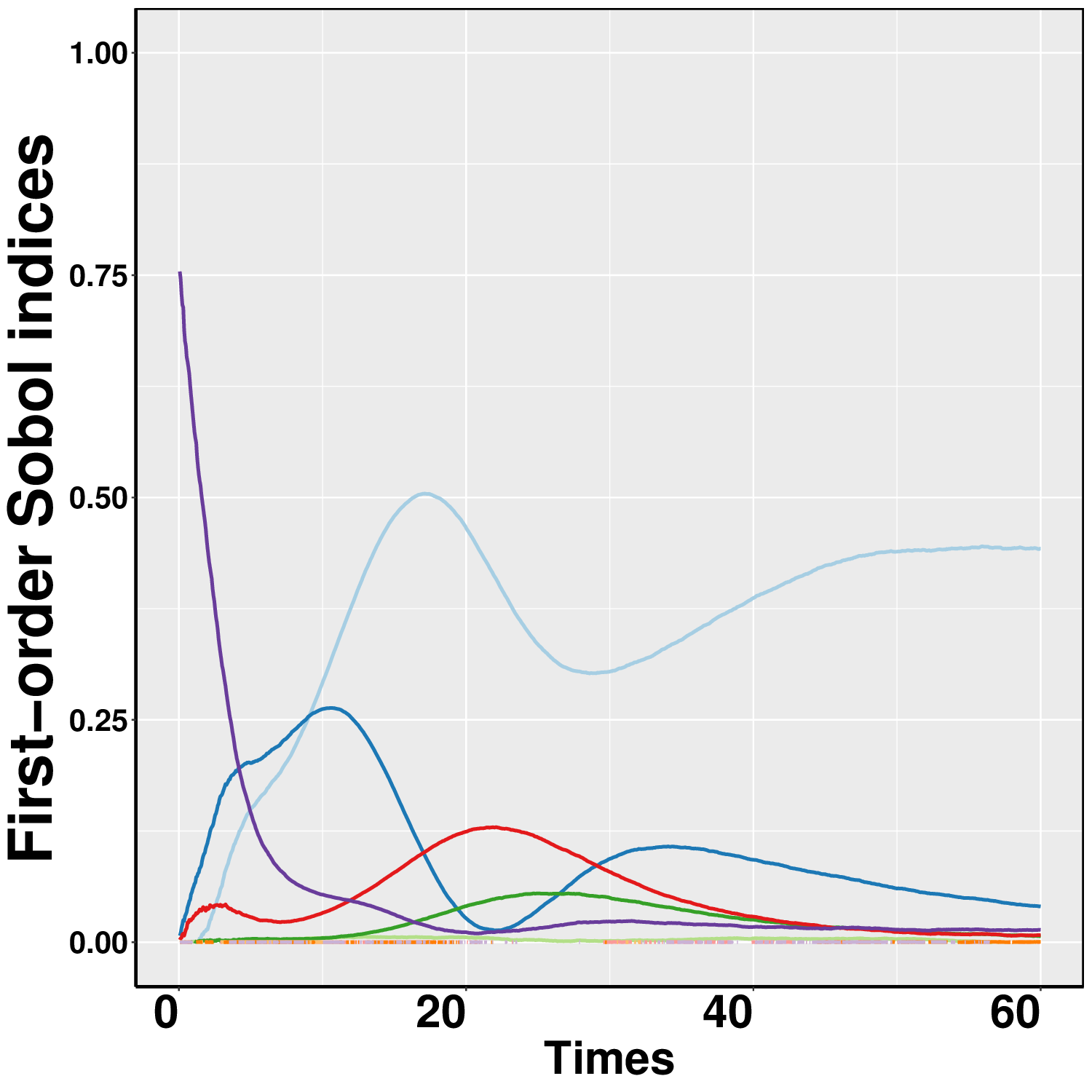} 
   &  
         \includegraphics[scale=0.22]{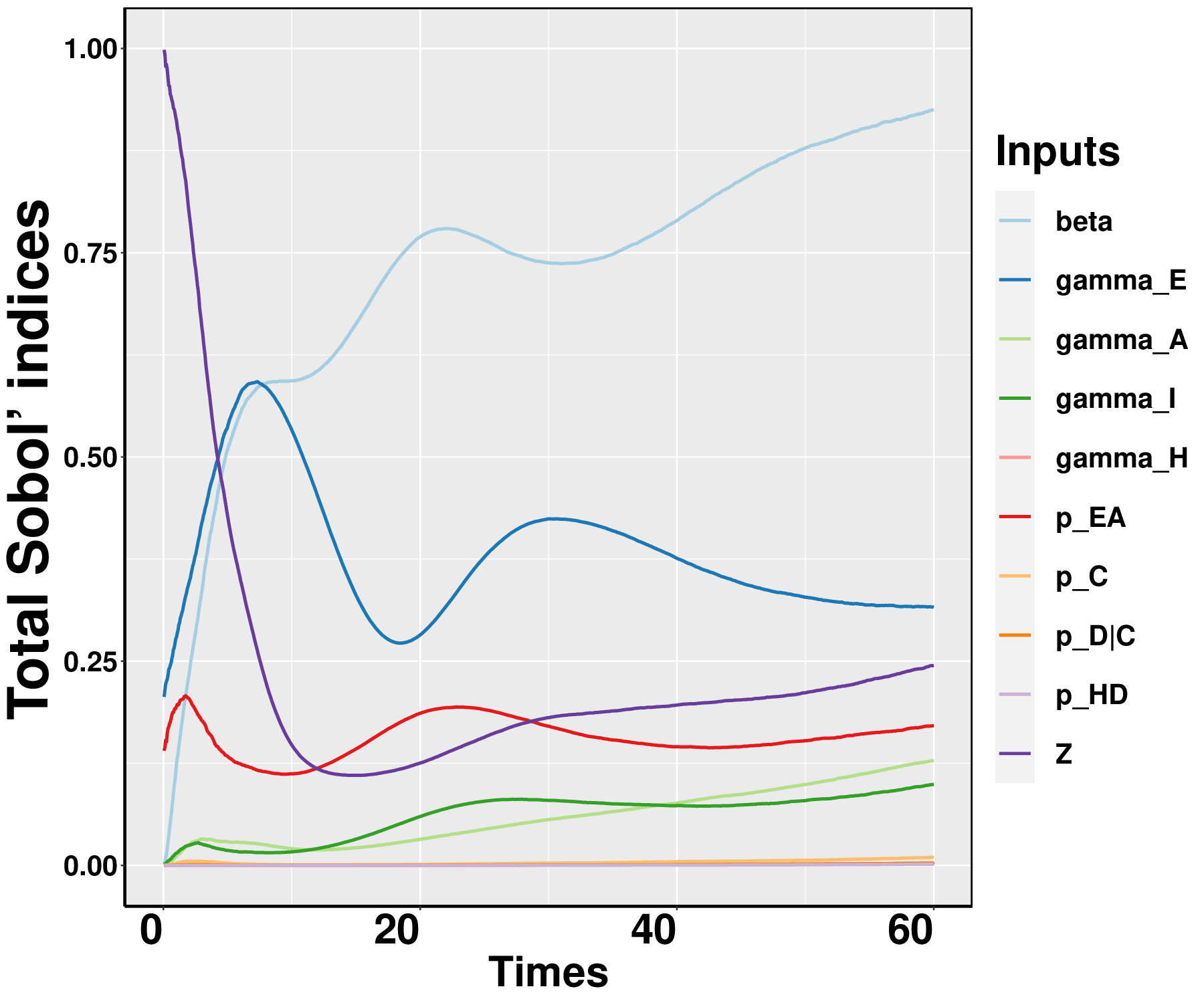}
                      \\
                       (e) & (f) \\
            \includegraphics[scale=0.22]{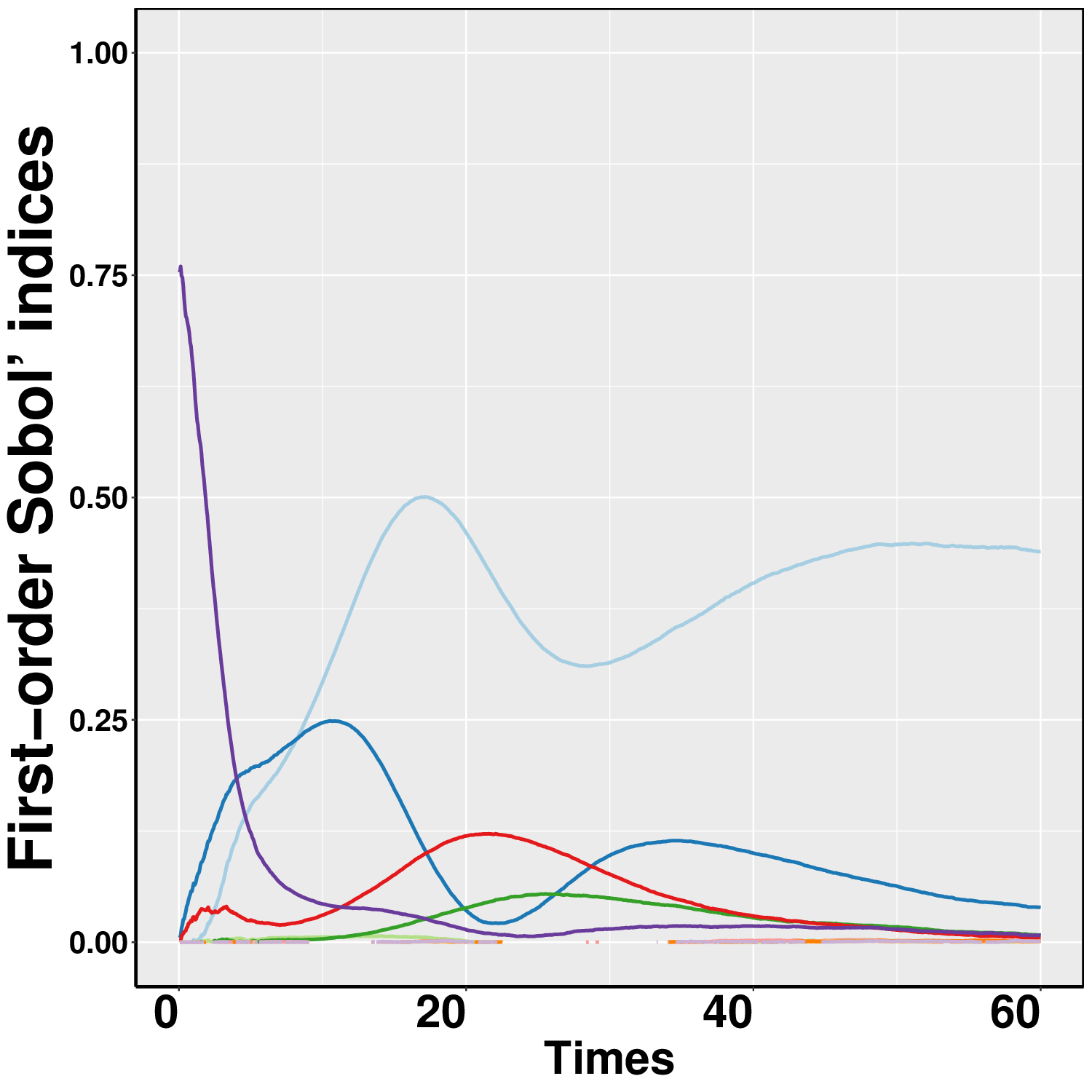} 
   &  
         \includegraphics[scale=0.22]{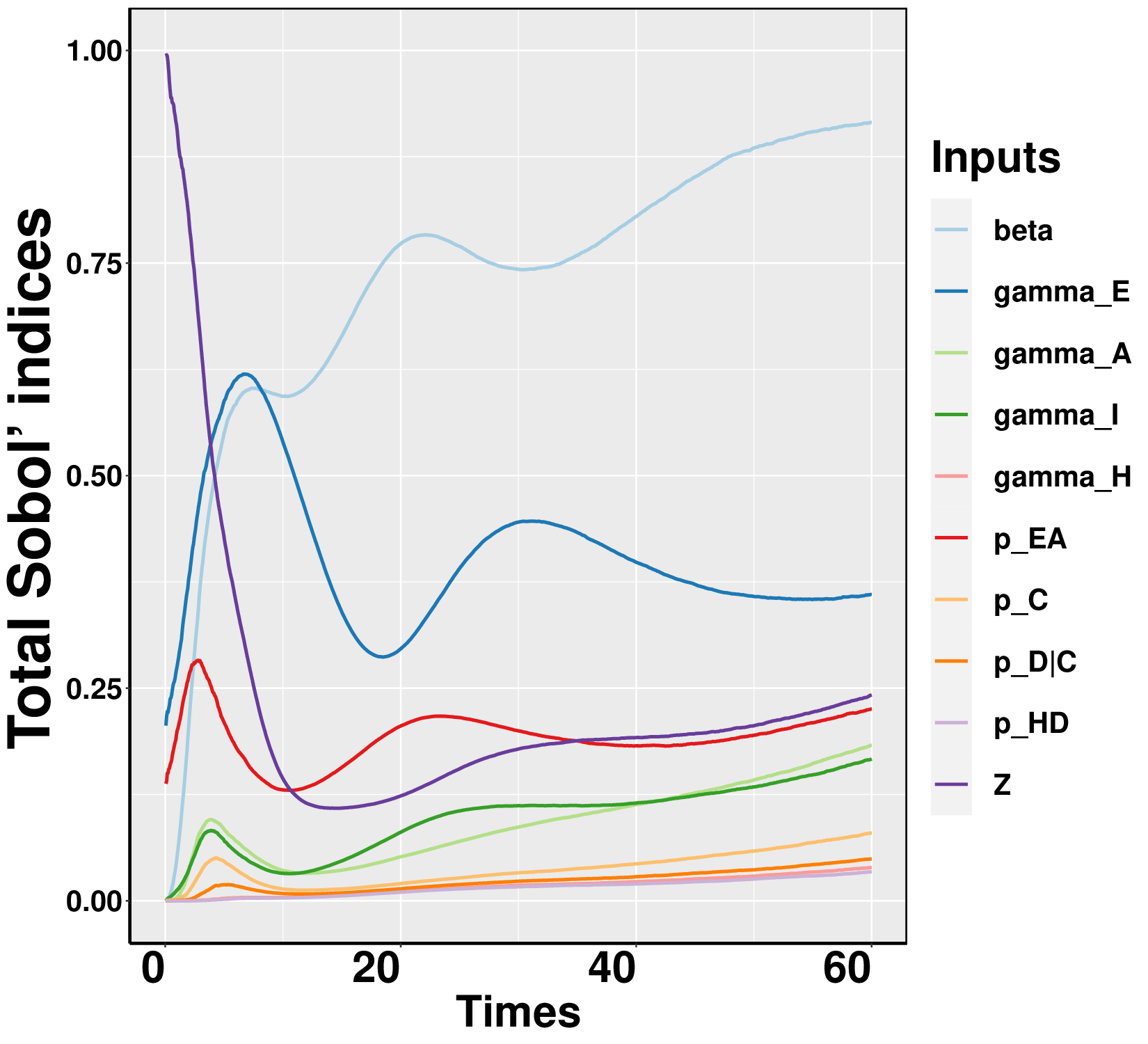}
   
     \end{tabular}
     \caption{Mean dynamical evolution of (left) first-order and
       (right) total Sobol' indices for $Y_I$ with respect to (Figures (a) and (b)) Gillespie Direct Method, (Figure (c) and (d)) Gillespie First Reaction, (Figures (e) and (f)) Modified Next Reaction algorithm. The
       mean is computed from $50$ independent repetitions of the
       estimation procedure performed with two independent
         designs of $n=2000$ input-output samples.}
        \label{dyn}
\end{figure}

The mean dynamical evolution of first-order and total Sobol' index
estimates is displayed on Figure~\ref{dyn}. Each mean is computed from
$50$ independent repetitions.  
At the beginning of the epidemic, the number of infected individuals
is mostly sensitive to $Z$---that is, to random
fluctuations inherent to the model. This confirms that intrinsic
randomness rules the dynamics in the emergence phase of an epidemic
disease. While the epidemic evolves, the main effect of $Z$ quickly
drops and some uncertain parameters---namely, $\beta$, $\gamma_E$ and to a less
extent $p_{EA}$---gain more influence. The uncertain parameter $\beta$, in
particular, becomes much more important than any other input and
remains so until the end. Notice that, except $\beta$, the main effect
of every input (both the uncertain parameters and intrinsic noise) approaches zero as the epidemic goes to its end, while
the opposite is true for total effects. This indicates that
interactions become more prevalent near the end of the epidemic.

\begin{figure}[H]
   \centering
   \includegraphics[width=\textwidth]{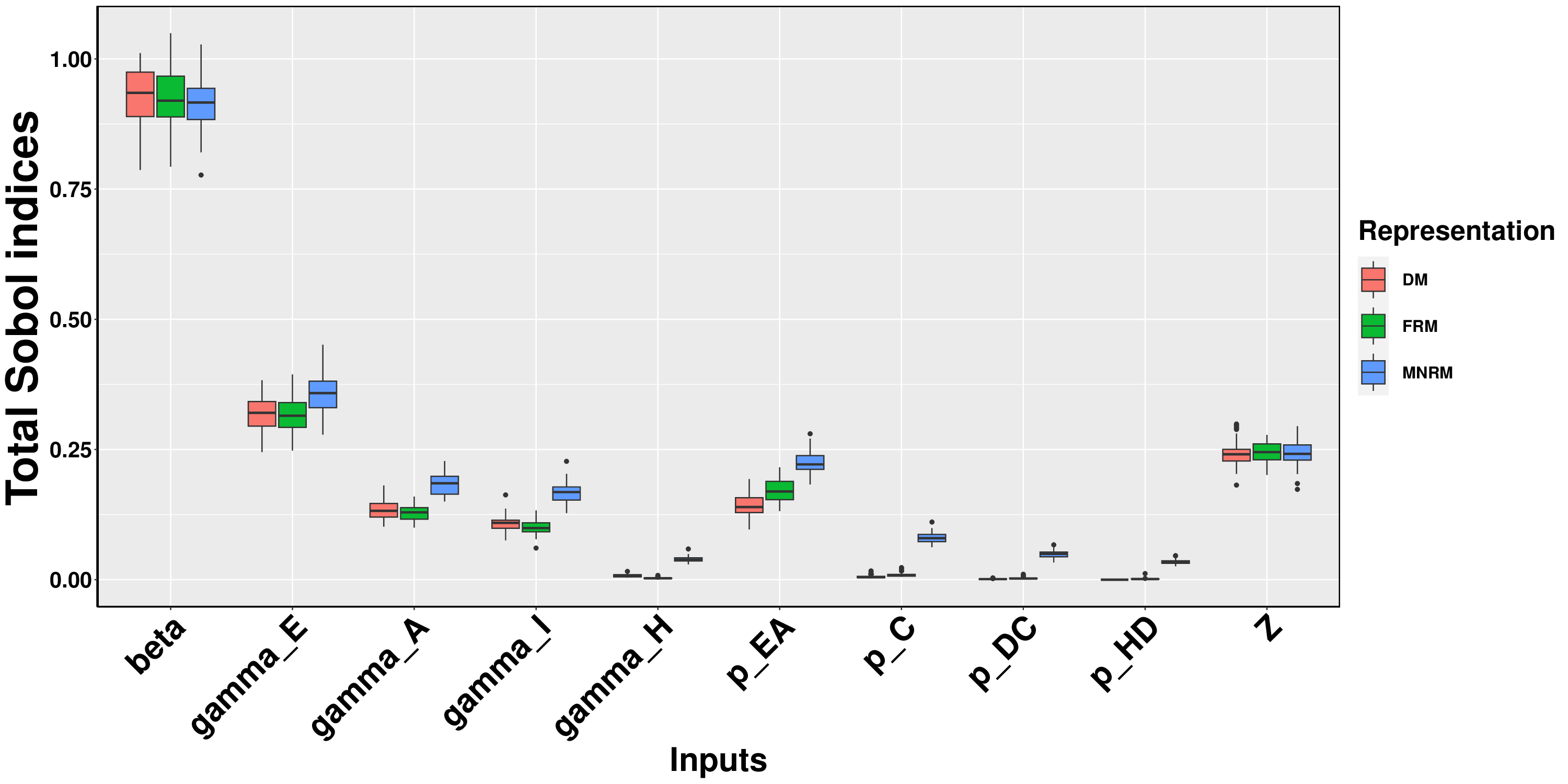}
   \caption{Zoom on total Sobol' indices at time point $t=60$.}
   \label{zoom}
\end{figure}

\begin{figure}[H]
\hspace{-1.2cm}
     \begin{subfigure}[b]{.3\textwidth}
        \includegraphics[scale=0.45]{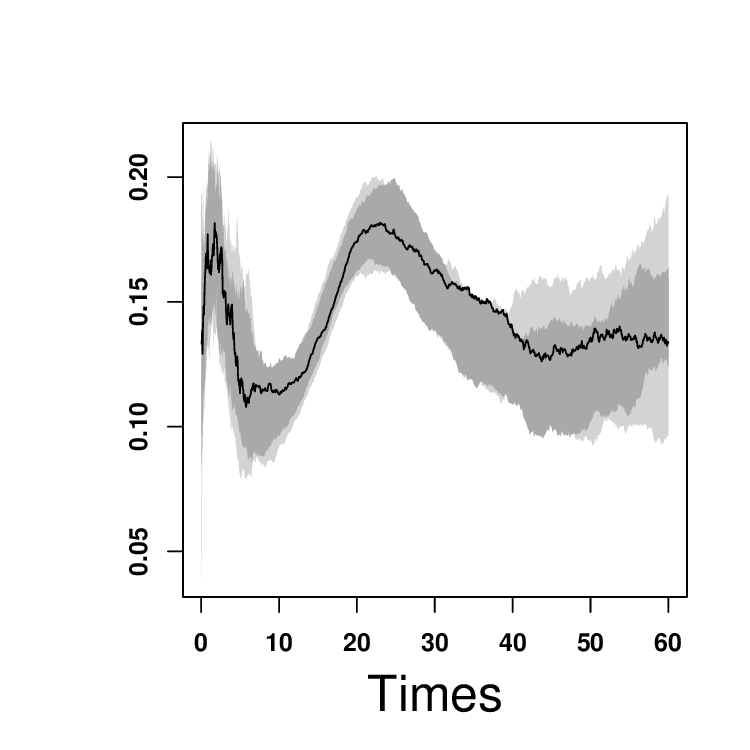} 
     \end{subfigure}
     \hspace{0.5cm}
     \begin{subfigure}[b]{.3\textwidth}
         \includegraphics[scale=0.45]{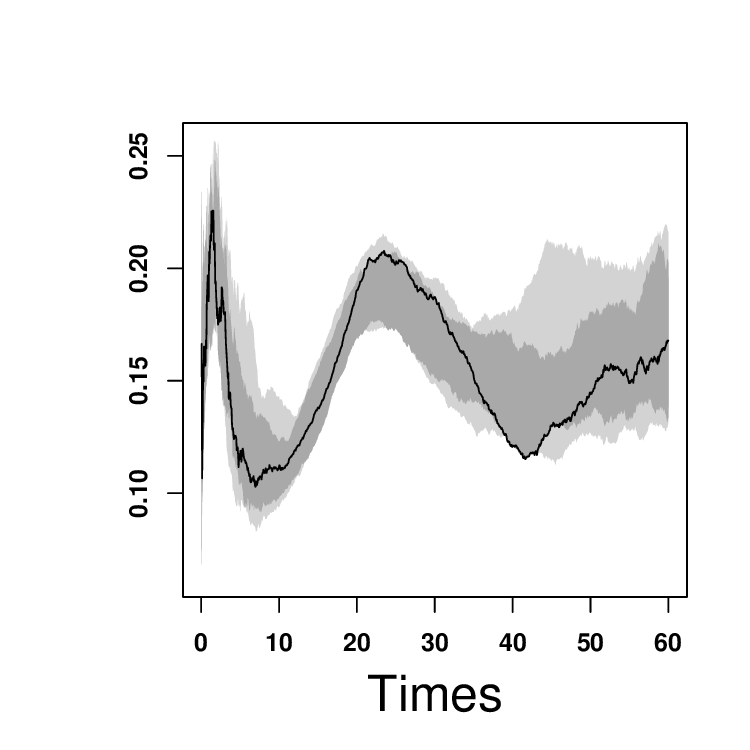}
     \end{subfigure}
     \hspace{0.5cm}
     \begin{subfigure}[b]{.3\textwidth}
         \includegraphics[scale=0.45]{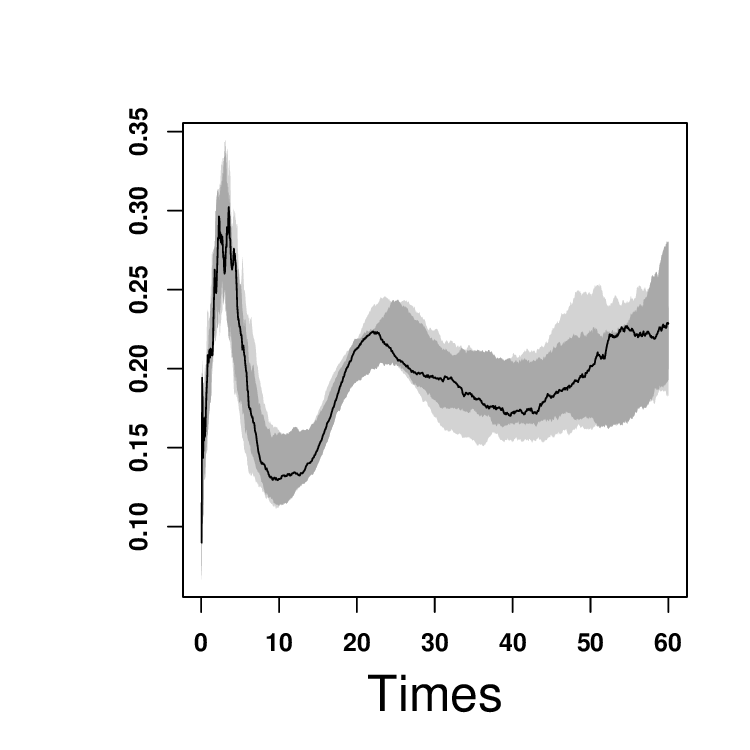}
     \end{subfigure}
       \caption{Functional HDR boxplots of dynamical total indices associated to parameter $p_{E,A}$, obtained from $50$ independent repetitions of the estimation procedure. (left) Gillespie Direct Method, (middle) Gillespie First Reaction Method, (right) Modified Next Reaction Method. The $50 \, \%$ HDR is plotted in dark gray, the $100 \, \%$ HDR in light gray and
the modal curve, that is the curve in the sample with the highest density is represented by a black solid line.}
\label{hdr}
\end{figure}

Although the most salient features of the performed sensitivity
analyses are shared between the three algorithms, we do observe some
differences in the mean dynamics across the three algorithms. These
differences seem to be significant: see Figure~\ref{hdr}, where the
sampling variability of the dynamics of first-order and total Sobol'
index estimates associated to $p_{EA}$ are displayed with functional
boxplots, namely highest density region (HDR) boxplots, obtained by
using the R package rainbow developed by
\cite{hyndman2010rainbow}. The HDR boxplot is a vizualization tool for
functional data based on kernel density estimation of the scores
associated to the two first principal components of the functional
data (see \cite{hyndman1996computing} for further details).  The
picture clearly indicates that the differences in the mean dynamics
obtained from the three different algorithms cannot be attributed to
sample variability alone.  As another example, a zoom in the time
$t=60$ (see Figure \ref{zoom}) shows significant differences for the
total index estimates of the parameters $\gamma_I$, $\gamma_H$, $p_C$,
$p_{DC}$ and $p_{HD}$, and to a less extent $\gamma_1$ and $p_{EA}$.


\subsection{Some thoughts about the choice of representations}\label{sec:choice}
The numerical experiments confirm that the sensitivity analysis results depend on the choice of the simulation algorithm. An interesting conclusion is that Gillespie algorithms are less prone to interactions between uncertain parameters and intrinsic randomness. It implies that simulations with Gillespie algorithms are more robust to a local perturbation of uncertain input parameters as we can see below by perturbing parameter $\beta$.



\begin{figure}[H]
\hspace{-1.2cm}
     \begin{subfigure}[b]{.3\textwidth}
        \includegraphics[scale=0.45]{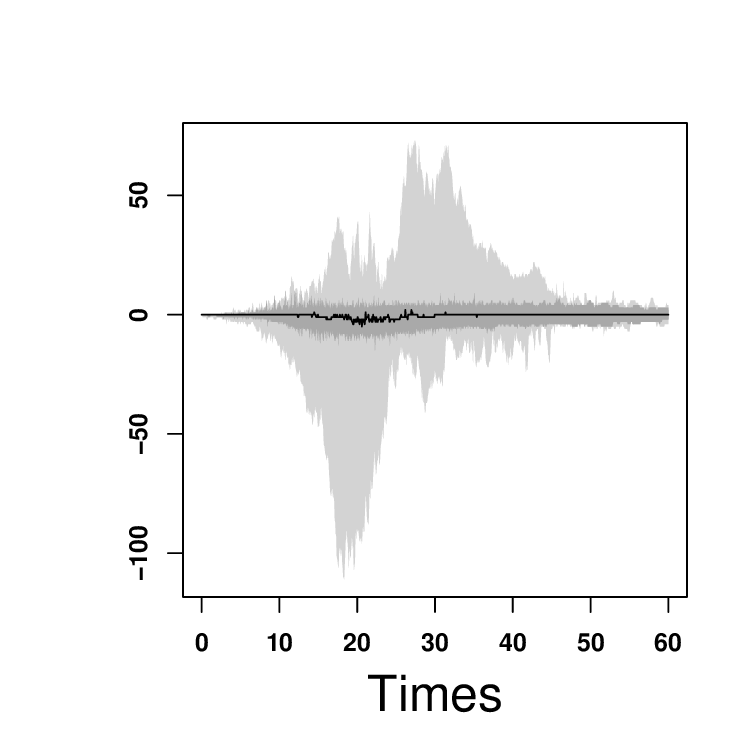} 
     \end{subfigure}
     \hspace{0.5cm}
     \begin{subfigure}[b]{.3\textwidth}
         \includegraphics[scale=0.45]{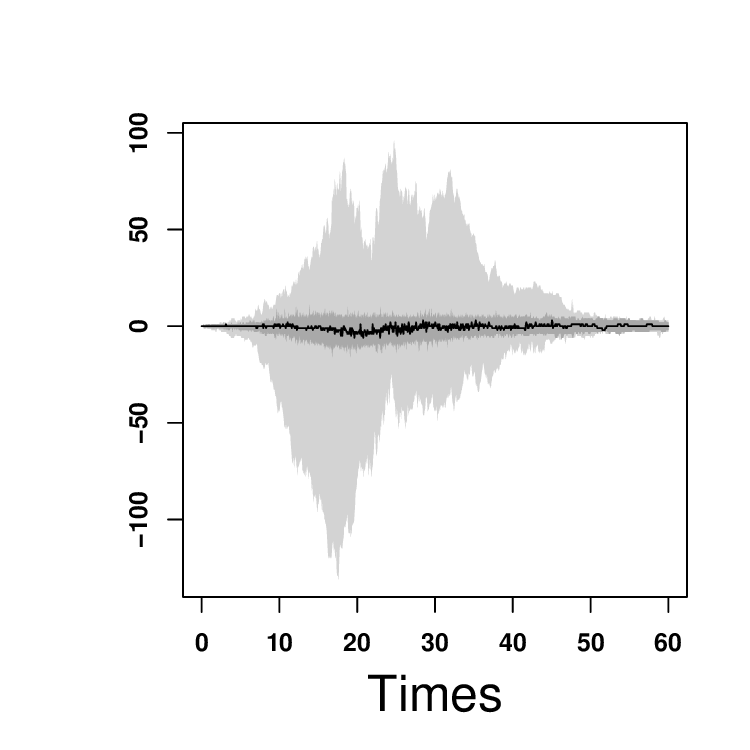}
     \end{subfigure}
     \hspace{0.5cm}
     \begin{subfigure}[b]{.3\textwidth}
         \includegraphics[scale=0.45]{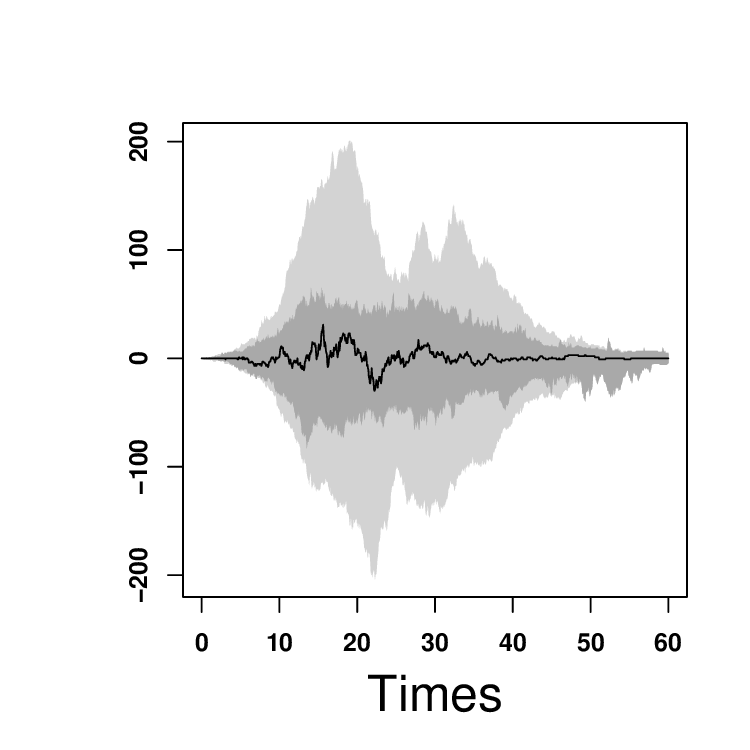}
     \end{subfigure}
       \caption{Functional HDR boxplots of differences of the dynamical number of symptomatic infectious individuals computed with uncertain parameters fixed to their nominal value and by perturbing parameter $\beta$ by 5\% from its nominal value: (left) Gillespie Direct Method, (middle) Gillespie First Reaction Method, (right) Modified Next Reaction Method. The $50 \, \%$ HDR is plotted in dark gray, the $100 \, \%$ HDR in light gray and
the modal curve, that is the curve in the sample with the highest density is represented by a black solid line. Functional HDR boxplots are plotted from $2 \, 000$ independent realizations.}
    \label{aggdiff}
\end{figure}

To plot Figure \ref{aggdiff}, we first simulate, for each simulation algorithm (Gillespie Direct Method, Gillespie First Reaction, Modified Next Reaction), $2\, 000$ trajectories (corresponding to $2 \, 000$ different seeds) of the difference between the number of symptomatic infectious individuals computed with all uncertain parameters fixed to their nominal value and the number of symptomatic infectious individuals computed by perturbing only parameter $\beta$ by $5 \, \%$ from its nominal value. \modif{Then in Figure \ref{aggdiff} are plotted highest density region (HDR) boxplots. It is clear on these plots that the small perturbation applied to parameter $\beta$ has a much stronger impact when Modified Next Reaction Method is used for simulations. We thus expect that quantities calculated from a Monte-Carlo sampling scheme are less robust to an inaccurate estimate of parameter $\beta$ when using simulations based on Modifed Next Reaction Algorithm.} 

\section{Conclusion}\label{conclu}
\modif{In this work, we leveraged three different exact simulation algorithms for continuous-time Markov chains from the state of the art which we combined with common tools from variance-based sensitivity analysis to perform a global sensitivity analysis of stochastic compartmental models. Such a methodology was introduced by \cite{Navarro} in the framework of chemical reaction networks, and using simulations from Modified Next Reaction Algorithm. In this paper, we discussed for the first time the impact of the choice of the algorithm used for model simulations on the result of global sensitivity analysis. We implemented and compared three sensitivity analyses based on simulations obtained from different exact simulation algorithms of a SARS-CoV-2 epidemic model. We observed that the different simulation algorithms are not equivalent in terms of robustness with respect to the uncertainty on epidemic parameters. Indeed with Figure \ref{aggdiff} we exhibited that variations in the value of parameter $\beta$ have a stronger influence on the variability of the simulated number of infectious individuals by using simulations produced with Modified Next Reaction algorithm.}

In the present paper, we considered Markovian models only. However an interesting follow-up would be to extend our results to non-Markovian stochastic processes by using Sellke's construction (\cite{sellke_1983}).

\modif{\section*{Acknowledgment}}

\modif{We thank the Associate Editor and two anonymous reviewers for their
thorough reading and constructive
feedback that led to an improved version of this manuscript.}

\section*{Declaration}

Declarations of interest: none.

\bibliographystyle{abbrvnat}
\bibliography{mabiblio.bib}

\begin{thebibliography}{47}
\providecommand{\natexlab}[1]{#1}
\providecommand{\url}[1]{\texttt{#1}}
\expandafter\ifx\csname urlstyle\endcsname\relax
  \providecommand{\doi}[1]{doi: #1}\else
  \providecommand{\doi}{doi: \begingroup \urlstyle{rm}\Url}\fi

\bibitem[Anderson(2007)]{anderson}
D.~F. Anderson.
\newblock A modified next reaction method for simulating chemical systems with
  time dependent propensities and delays.
\newblock \emph{The Journal of Chemical Physics}, 127\penalty0 (21):\penalty0
  214107, 2007.
\newblock \doi{10.1063/1.2799998}.

\bibitem[Bittihn and Golestanian(2020)]{bittihn_stochastic_2020}
P.~Bittihn and R.~Golestanian.
\newblock Stochastic effects on the dynamics of an epidemic due to population
  subdivision.
\newblock \emph{Chaos: An Interdisciplinary Journal of Nonlinear Science},
  30\penalty0 (10):\penalty0 101--102, Oct. 2020.
\newblock ISSN 1054-1500.
\newblock \doi{10.1063/5.0028972}.

\bibitem[Brauer(2008)]{Brauer2008}
F.~Brauer.
\newblock \emph{Compartmental Models in Epidemiology}, pages 19--79.
\newblock Springer Berlin Heidelberg, Berlin, Heidelberg, 2008.
\newblock ISBN 978-3-540-78911-6.
\newblock \doi{10.1007/978-3-540-78911-6_2}.

\bibitem[Britton(2009)]{Britton2009StochasticEM}
T.~Britton.
\newblock Stochastic epidemic models: a survey.
\newblock \emph{Mathematical biosciences}, 225 1:\penalty0 24--35, 2009.

\bibitem[Cazelles et~al.(2021)Cazelles, Champagne, Nguyen-Van-Yen, Comiskey,
  Vergu, and Roche]{cazelles}
B.~Cazelles, C.~Champagne, B.~Nguyen-Van-Yen, C.~Comiskey, E.~Vergu, and
  B.~Roche.
\newblock A mechanistic and data-driven reconstruction of the time-varying
  reproduction number: Application to the covid-19 epidemic.
\newblock \emph{PLOS Computational Biology}, 17\penalty0 (7):\penalty0 1--20,
  07 2021.
\newblock \doi{10.1371/journal.pcbi.1009211}.

\bibitem[Courcoul et~al.(2011)Courcoul, Monod, Nielen, Klinkenberg, Hogerwerf,
  Beaudeau, and Vergu]{courcoul}
A.~Courcoul, H.~Monod, M.~Nielen, D.~Klinkenberg, L.~Hogerwerf, F.~Beaudeau,
  and E.~Vergu.
\newblock Modelling the effect of heterogeneity of shedding on the within herd
  coxiella burnetii spread and identification of key parameters by sensitivity
  analysis.
\newblock \emph{Journal of Theoretical Biology}, 284\penalty0 (1):\penalty0
  130--141, 2011.
\newblock ISSN 0022-5193.
\newblock \doi{https://doi.org/10.1016/j.jtbi.2011.06.017}.

\bibitem[Cristancho~Fajardo et~al.(2021)Cristancho~Fajardo, Ezanno, and
  Vergu]{Cristancho}
L.~Cristancho~Fajardo, P.~Ezanno, and E.~Vergu.
\newblock Accounting for farmers’ control decisions in a model of pathogen
  spread through animal trade.
\newblock \emph{Scientific Reports}, 11\penalty0 (1):\penalty0 9581, May 2021.
\newblock ISSN 2045-2322.
\newblock \doi{10.1038/s41598-021-88471-6}.

\bibitem[Da~Veiga et~al.(2021)Da~Veiga, Gamboa, Iooss, and
  Prieur]{da2021basics}
S.~Da~Veiga, F.~Gamboa, B.~Iooss, and C.~Prieur.
\newblock \emph{Basics and trends in sensitivity analysis: Theory and practice
  in R}.
\newblock SIAM, 2021.

\bibitem[Dupuy et~al.(2015)Dupuy, Helbert, and Franco]{dicedesign}
D.~Dupuy, C.~Helbert, and J.~Franco.
\newblock {DiceDesign} and {DiceEval}: Two {R} packages for design and analysis
  of computer experiments.
\newblock \emph{Journal of Statistical Software}, 65\penalty0 (11):\penalty0
  1--38, 2015.

\bibitem[Ethier and Kurtz(1986)]{Kurtz1}
S.~N. Ethier and T.~G. Kurtz.
\newblock \emph{Markov processes -- characterization and convergence}, chapter
  4, 6.
\newblock Wiley Series in Probability and Mathematical Statistics: Probability
  and Mathematical Statistics. John Wiley \& Sons Inc., New York, 1986.
\newblock ISBN 0-471-08186-8.

\bibitem[\'Etoré et~al.(2020)\'Etoré, Prieur, Pham, and Li]{etore}
P.~\'Etoré, C.~Prieur, D.~K. Pham, and L.~Li.
\newblock Global sensitivity analysis for models described by stochastic
  differential equations.
\newblock \emph{Methodology and Computing in Applied Probability}, 22\penalty0
  (2):\penalty0 803--831, June 2020.
\newblock ISSN 1387-5841, 1573-7713.
\newblock \doi{10.1007/s11009-019-09732-6}.

\bibitem[Fort et~al.(2021)Fort, Klein, and Lagnoux]{fort2020global}
J.-C. Fort, T.~Klein, and A.~Lagnoux.
\newblock Global sensitivity analysis and {Wasserstein} spaces.
\newblock \emph{SIAM/ASA Journal on Uncertainty Quantification}, 9\penalty0
  (2):\penalty0 880--921, 2021.
\newblock \doi{10.1137/20M1354957}.

\bibitem[Gamboa et~al.(2014)Gamboa, Janon, Klein, and Lagnoux]{gamboa2}
F.~Gamboa, A.~Janon, T.~Klein, and A.~Lagnoux.
\newblock {Sensitivity analysis for multidimensional and functional outputs}.
\newblock \emph{Electronic Journal of Statistics}, 8\penalty0 (1):\penalty0 575
  -- 603, 2014.
\newblock \doi{10.1214/14-EJS895}.

\bibitem[Gillespie(1976)]{GILLESPIE}
D.~T. Gillespie.
\newblock A general method for numerically simulating the stochastic time
  evolution of coupled chemical reactions.
\newblock \emph{Journal of Computational Physics}, 22\penalty0 (4):\penalty0
  403--434, 1976.
\newblock ISSN 0021-9991.
\newblock \doi{https://doi.org/10.1016/0021-9991(76)90041-3}.

\bibitem[Goel et~al.(2023)Goel, Bhatia, Tripathi, Bugalia, Rana, and
  Bajiya]{goel2023sirc}
S.~Goel, S.~K. Bhatia, J.~P. Tripathi, S.~Bugalia, M.~Rana, and V.~P. Bajiya.
\newblock S{IRC} epidemic model with cross-immunity and multiple time delays.
\newblock \emph{Journal of Mathematical Biology}, 87\penalty0 (3):\penalty0 42,
  2023.

\bibitem[Hanthanan~Arachchilage et~al.(2023)Hanthanan~Arachchilage, Hussaini,
  Cogan, and Cortez]{hanthanan2023exploring}
K.~Hanthanan~Arachchilage, M.~Y. Hussaini, N.~Cogan, and M.~H. Cortez.
\newblock Exploring how ecological and epidemiological processes shape
  multi-host disease dynamics using global sensitivity analysis.
\newblock \emph{Journal of Mathematical Biology}, 86\penalty0 (5):\penalty0 83,
  2023.

\bibitem[Hart et~al.(2017)Hart, Alexanderian, and Gremaud]{Hart}
J.~L. Hart, A.~Alexanderian, and P.~A. Gremaud.
\newblock Efficient computation of {Sobol'} indices for stochastic models.
\newblock \emph{SIAM Journal on Scientific Computing}, 39\penalty0
  (4):\penalty0 A1514--A1530, 2017.
\newblock \doi{10.1137/16M106193X}.

\bibitem[Homma and Saltelli(1996)]{HOMMA}
T.~Homma and A.~Saltelli.
\newblock Importance measures in global sensitivity analysis of nonlinear
  models.
\newblock \emph{Reliability Engineering and System Safety}, 52\penalty0
  (1):\penalty0 1--17, 1996.
\newblock ISSN 0951-8320.
\newblock \doi{https://doi.org/10.1016/0951-8320(96)00002-6}.

\bibitem[Hyndman(1996)]{hyndman1996computing}
R.~J. Hyndman.
\newblock Computing and graphing highest density regions.
\newblock \emph{The American Statistician}, 50\penalty0 (2):\penalty0 120--126,
  1996.

\bibitem[Hyndman and Shang(2010)]{hyndman2010rainbow}
R.~J. Hyndman and H.~L. Shang.
\newblock Rainbow plots, bagplots, and boxplots for functional data.
\newblock \emph{Journal of Computational and Graphical Statistics}, 19\penalty0
  (1):\penalty0 29--45, 2010.

\bibitem[Iooss et~al.(2021)Iooss, {Da~Veiga}, Janon, and Pujol]{sensitivity}
B.~Iooss, S.~{Da~Veiga}, A.~Janon, and G.~Pujol.
\newblock \emph{sensitivity: Global Sensitivity Analysis of Model Outputs},
  2021.
\newblock URL \url{https://CRAN.R-project.org/package=sensitivity}.
\newblock R package version 1.24.0.

\bibitem[Jimenez et~al.(2017)Jimenez, Le~Maître, and Knio]{jimenez}
M.~N. Jimenez, O.~P. Le~Maître, and O.~M. Knio.
\newblock Nonintrusive polynomial chaos expansions for sensitivity analysis in
  stochastic differential equations.
\newblock \emph{SIAM/ASA Journal on Uncertainty Quantification}, 5\penalty0
  (1):\penalty0 378--402, Jan. 2017.
\newblock ISSN 2166-2525.
\newblock \doi{10.1137/16M1061989}.

\bibitem[Karlin and Taylor(1981)]{KarlinTaylorSecondCourse}
S.~Karlin and H.~M. Taylor.
\newblock \emph{A second course on stochastic processes}.
\newblock Academic Press, 1981.

\bibitem[Knock et~al.(2021)Knock, Whittles, Lees, Perez-Guzman, Verity,
  FitzJohn, Gaythorpe, Imai, Hinsley, Okell, Rosello, Kantas, Walters, Bhatia,
  Watson, Whittaker, Cattarino, Boonyasiri, Djaafara, Fraser, Fu, Wang, Xi,
  Donnelly, Jauneikaite, Laydon, White, Ghani, Ferguson, Cori, and
  Baguelin]{Knock2021}
E.~S. Knock, L.~K. Whittles, J.~A. Lees, P.~N. Perez-Guzman, R.~Verity, R.~G.
  FitzJohn, K.~A.~M. Gaythorpe, N.~Imai, W.~Hinsley, L.~C. Okell, A.~Rosello,
  N.~Kantas, C.~E. Walters, S.~Bhatia, O.~J. Watson, C.~Whittaker,
  L.~Cattarino, A.~Boonyasiri, B.~A. Djaafara, K.~Fraser, H.~Fu, H.~Wang,
  X.~Xi, C.~A. Donnelly, E.~Jauneikaite, D.~J. Laydon, P.~J. White, A.~C.
  Ghani, N.~M. Ferguson, A.~Cori, and M.~Baguelin.
\newblock Key epidemiological drivers and impact of interventions in the 2020
  sars-cov-2 epidemic in england.
\newblock \emph{Science Translational Medicine}, 13\penalty0 (602):\penalty0
  42--62, 2021.
\newblock \doi{10.1126/scitranslmed.abg4262}.

\bibitem[Kurtz(1982)]{Kurtz2}
T.~G. Kurtz.
\newblock Representation and approximation of counting processes.
\newblock In W.~H. Fleming and L.~G. Gorostiza, editors, \emph{Advances in
  Filtering and Optimal Stochastic Control}, pages 177--191, Berlin,
  Heidelberg, 1982. Springer Berlin Heidelberg.
\newblock ISBN 978-3-540-39517-1.

\bibitem[Lamboni et~al.(2011)Lamboni, Hervé, and David]{lamboni}
M.~Lamboni, M.~Hervé, and M.~David.
\newblock Multivariate sensitivity analysis to measure global contribution of
  input factors in dynamic models.
\newblock \emph{Reliability Engineering and System Safety}, 96\penalty0
  (4):\penalty0 450 -- 459, 2011.
\newblock ISSN 0951-8320.
\newblock \doi{https://doi.org/10.1016/j.ress.2010.12.002}.

\bibitem[{Le Maître} and Knio(2015)]{LEMAITRE2015107}
O.~{Le Maître} and O.~Knio.
\newblock {PC} analysis of stochastic differential equations driven by wiener
  noise.
\newblock \emph{Reliability Engineering \& System Safety}, 135:\penalty0
  107--124, 2015.
\newblock ISSN 0951-8320.
\newblock \doi{https://doi.org/10.1016/j.ress.2014.11.002}.

\bibitem[Le~Maître et~al.(2015)Le~Maître, Knio, and Moraes]{lemaitre}
O.~P. Le~Maître, O.~M. Knio, and A.~Moraes.
\newblock Variance decomposition in stochastic simulators.
\newblock \emph{The Journal of Chemical Physics}, 142\penalty0 (24):\penalty0
  244115, 2015.
\newblock \doi{10.1063/1.4922922}.

\bibitem[Lin and Tang(2015)]{lintan15}
C.~Lin and B.~Tang.
\newblock Latin hypercubes and space-filling designs.
\newblock In A.~Dean, M.~Morris, J.~Stufken, and D.~Bingham, editors,
  \emph{Handbook of design and analysis of experiments}, Handbooks of modern
  statistical methods. Chapman \& Hall/CRC, 2015.

\bibitem[Marino et~al.(2008)Marino, Hogue, Ray, and Kirschner]{MARINO2008178}
S.~Marino, I.~B. Hogue, C.~J. Ray, and D.~E. Kirschner.
\newblock A methodology for performing global uncertainty and sensitivity
  analysis in systems biology.
\newblock \emph{Journal of Theoretical Biology}, 254\penalty0 (1):\penalty0
  178--196, 2008.
\newblock ISSN 0022-5193.
\newblock \doi{https://doi.org/10.1016/j.jtbi.2008.04.011}.
\newblock URL
  \url{https://www.sciencedirect.com/science/article/pii/S0022519308001896}.

\bibitem[Marrel et~al.(2012)Marrel, Iooss, Da~Veiga, and
  Ribatet]{marrel_global_2012}
A.~Marrel, B.~Iooss, S.~Da~Veiga, and M.~Ribatet.
\newblock Global sensitivity analysis of stochastic computer models with joint
  metamodels.
\newblock \emph{Statistics and Computing}, 22\penalty0 (3):\penalty0 833--847,
  May 2012.
\newblock ISSN 1573-1375.
\newblock \doi{10.1007/s11222-011-9274-8}.

\bibitem[Massard et~al.(2022)Massard, Eftimie, Perasso, and
  Saussereau]{MASSARD2022111117}
M.~Massard, R.~Eftimie, A.~Perasso, and B.~Saussereau.
\newblock A multi-strain epidemic model for covid-19 with infected and
  asymptomatic cases: Application to french data.
\newblock \emph{Journal of Theoretical Biology}, 545:\penalty0 111117, 2022.
\newblock ISSN 0022-5193.
\newblock \doi{https://doi.org/10.1016/j.jtbi.2022.111117}.
\newblock URL
  \url{https://www.sciencedirect.com/science/article/pii/S0022519322001151}.

\bibitem[Mazo(2021)]{mazo}
G.~Mazo.
\newblock A trade-off between explorations and repetitions for estimators of
  two global sensitivity indices in stochastic models induced by probability
  measures.
\newblock \emph{SIAM/ASA Journal on Uncertainty Quantification}, 9\penalty0
  (4):\penalty0 1673--1713, 2021.
\newblock \doi{10.1137/19M1272706}.

\bibitem[Navarro~Jimenez et~al.(2016)Navarro~Jimenez, Le~Maître, and
  Knio]{Navarro}
M.~Navarro~Jimenez, O.~P. Le~Maître, and O.~M. Knio.
\newblock Global sensitivity analysis in stochastic simulators of uncertain
  reaction networks.
\newblock \emph{The Journal of Chemical Physics}, 145\penalty0 (24):\penalty0
  244106, 2016.
\newblock \doi{10.1063/1.4971797}.

\bibitem[Ngonghala et~al.(2015)Ngonghala, Teboh-Ewungkem, and
  Ngwa]{ngonghala2015persistent}
C.~N. Ngonghala, M.~I. Teboh-Ewungkem, and G.~A. Ngwa.
\newblock Persistent oscillations and backward bifurcation in a malaria model
  with varying human and mosquito populations: implications for control.
\newblock \emph{Journal of mathematical biology}, 70\penalty0 (7):\penalty0
  1581--1622, 2015.

\bibitem[{R Core Team}(2021)]{Rsoft}
{R Core Team}.
\newblock \emph{R: A Language and Environment for Statistical Computing}.
\newblock R Foundation for Statistical Computing, Vienna, Austria, 2021.
\newblock URL \url{https://www.R-project.org/}.

\bibitem[Richard et~al.(2021)Richard, Alizon, Choisy, Sofonea, and
  Djidjou-Demasse]{Richard}
Q.~Richard, S.~Alizon, M.~Choisy, M.~T. Sofonea, and R.~Djidjou-Demasse.
\newblock Age-structured non-pharmaceutical interventions for optimal control
  of covid-19 epidemic.
\newblock \emph{PLOS Computational Biology}, 17\penalty0 (3):\penalty0 1--25,
  03 2021.
\newblock \doi{10.1371/journal.pcbi.1008776}.

\bibitem[Rimbaud et~al.(2018)Rimbaud, Bruchou, Dallot, Pleydell, Jacquot,
  Soubeyrand, and Thébaud]{Rimbaud}
L.~Rimbaud, C.~Bruchou, S.~Dallot, D.~R.~J. Pleydell, E.~Jacquot,
  S.~Soubeyrand, and G.~Thébaud.
\newblock Using sensitivity analysis to identify key factors for the
  propagation of a plant epidemic.
\newblock \emph{Royal Society Open Science}, 5\penalty0 (1):\penalty0 171435,
  2018.
\newblock \doi{10.1098/rsos.171435}.

\bibitem[Saltelli et~al.(2000)Saltelli, Chan, and Scott]{SAL2000}
A.~Saltelli, K.~Chan, and E.~M. Scott.
\newblock \emph{Sensitivity Analysis}.
\newblock John Wiley \& Sons, 2000.

\bibitem[Sellke(1983)]{sellke_1983}
T.~Sellke.
\newblock On the asymptotic distribution of the size of a stochastic epidemic.
\newblock \emph{Journal of Applied Probability}, 20\penalty0 (2):\penalty0
  390–394, 1983.
\newblock \doi{10.2307/3213811}.

\bibitem[Sobol'(1993)]{sobol}
I.~M. Sobol'.
\newblock Sensitivity analysis for non-linear mathematical models.
\newblock \emph{Mathematical Modelling and Computational Experiment},
  1:\penalty0 407--414, 1993.

\bibitem[Torii et~al.(2023)Torii, Begnini, Kroetz, Matar, Lopez, and
  Miguel]{torii2023global}
A.~J. Torii, R.~Begnini, H.~M. Kroetz, O.~M.~I. Matar, R.~H. Lopez, and
  L.~F.~F. Miguel.
\newblock Global sensitivity analysis for mathematical models comparison.
\newblock \emph{Computational and Applied Mathematics}, 42\penalty0
  (8):\penalty0 345, 2023.

\bibitem[Veiga(2021)]{daveiga}
S.~D. Veiga.
\newblock Kernel-based {ANOVA} decomposition and shapley effects -- application
  to global sensitivity analysis, 2021.
\newblock arXiv:2101.05487.

\bibitem[Williams and Rasmussen(2006)]{williams2006gaussian}
C.~K. Williams and C.~E. Rasmussen.
\newblock \emph{Gaussian processes for machine learning}, volume~2.
\newblock MIT press Cambridge, MA, 2006.

\bibitem[Yang et~al.(2016)Yang, Chen, and Xu]{yang2016effect}
J.~Yang, Y.~Chen, and F.~Xu.
\newblock Effect of infection age on an {SIS} epidemic model on complex
  networks.
\newblock \emph{Journal of mathematical biology}, 73:\penalty0 1227--1249,
  2016.

\bibitem[Zhu and Sudret(2021)]{zhu}
X.~Zhu and B.~Sudret.
\newblock Global sensitivity analysis for stochastic simulators based on
  generalized lambda surrogate models.
\newblock \emph{Reliability Engineering \& System Safety}, 214:\penalty0
  107815, 2021.
\newblock ISSN 0951-8320.
\newblock \doi{https://doi.org/10.1016/j.ress.2021.107815}.

\bibitem[Zhu and Sudret(2023)]{Zhu2023}
X.~Zhu and B.~Sudret.
\newblock Stochastic polynomial chaos expansions to emulate stochastic
  simulators.
\newblock \emph{International Journal for Uncertainty Quantification},
  13\penalty0 (2):\penalty0 31--52, 2023.
\newblock ISSN 2152-5080.

\end{thebibliography}
\end{document}